\documentclass[reprint,prb,amsmath,amssymb,aps,]{revtex4-2}

\usepackage{graphicx}
\usepackage{dcolumn}
\usepackage{bm}
\usepackage[colorlinks, allcolors=blue]{hyperref}
\usepackage[all]{hypcap}
\usepackage{amsmath}
\usepackage{caption}
\usepackage{subcaption}
\usepackage{multirow}
\usepackage{algorithm}
\usepackage{algpseudocode}
\usepackage[export]{adjustbox}
\usepackage{comment}
\usepackage{subcaption}
\usepackage{booktabs}
\newcommand{\ket}[1]{\vert #1\rangle}
\newcommand{\bra}[1]{\langle #1\vert}
\newcommand{\expect}[3]{\langle #1\vert #2 \vert #3 \rangle}
\newcommand{\braket}[2]{\langle #1\vert #2 \rangle}
\DeclareMathOperator{\Tr}{Tr}

\newcommand{\hide}[1]{}
\newcommand{\pref}[2]{\hyperref[#1]{\ref{#1}(#2)}}

\begin{document}
\title{Non-equilibrium Quantum Monte Carlo Algorithm for Stabilizer R\'enyi Entropy in Spin Systems}

\def\urbana{
The Anthony J. Leggett Institute for Condensed Matter Theory and IQUIST and NCSA Center for Artificial Intelligence Innovation and Department of Physics, University of Illinois at Urbana-Champaign, IL 61801, USA}

\author{Zejun Liu} 
\email{zejunliu@illinois.edu}
\affiliation{\urbana}
\author{Bryan K. Clark}
\email{bkclark@illinois.edu}
\affiliation{\urbana}
\date{\today}
\begin{abstract}
Quantum magic, or nonstabilizerness, provides a crucial characterization of quantum systems, regarding the classical simulability with stabilizer states. In this work, we propose a novel and efficient algorithm for computing  stabilizer R\'enyi entropy, one of the measures for quantum magic, in spin systems with sign-problem free Hamiltonians. This algorithm is based on the quantum Monte Carlo simulation of the path integral of the work between two partition function ensembles and it applies to all spatial dimensions and temperatures. We demonstrate this algorithm on the one and two dimensional transverse field Ising model at both finite and zero temperatures and show the quantitative agreements with tensor-network based algorithms. We analyze the computational cost and provide analytical and numerical evidences for it to be polynomial in system size. This work also suggests a unifying framework for calculating various types of entropy quantities including entanglement R\'enyi entropy and entanglement R\'enyi negativity. 
\end{abstract}
\maketitle
\section{Introduction}
How quantum a physical state is characterizes the potential of using that state for beyond classical behavior - e.g. using this state as a resource to solve classically intractable tasks~\cite{feynman2018simulating,preskill2018quantum}. In the past decades, one such measure has been quantum entanglement which has been intensively studied in various regimes such as quantum information, condensed matter, and high energy physics~\cite{Calabrese2004,Levin2006,Kitaev2006,Fradkin2006,Casini2007,Li2019,Chan2019,Skinner2019,faulkner2022snowmass,headrick2019lectures}. Product states, which have zero entanglement, are essentially classical and simulations where a pure state is always weakly entangled over all bipartitions can be implemented classically.   Quantum entanglement, though, is not the only measure of the ``quantumness'' of a state.  For example, stabilizer states can be highly entangled but they can be represented in a classically efficient way~\cite{Aaronson2004,gottesman1998heisenberg}.  The  nonstabilizerness of a state is called quantum magic and  characterizes the deviation of a quantum state from a stabilizer state, or alternatively, how many stabilizer states are required for an accurate classical description of this state~\cite{Bravyi2019simulationofquantum,Pashayan2022,oliviero2022measuring,Haug2023_PRXQ,Liu2022,Ellison2021symmetryprotected,Fliss2021}.

Like quantum entanglement, there are many different ways to quantify quantum magic, with the stabilizer R\'enyi entropy (SRE) ~\cite{Leone2022} having a number of desirable properties including being strictly non-negative and only zero when the states are stabilizer states; invariant under transformation of any Clifford operators; and additive when the states are separable~\cite{Leone2022}.  SRE quantifies the uniformity of the decomposition of a quantum state into Pauli strings.  Fully uniform states, whose SRE is zero, can be classically simulated with a Clifford circuit (with measured ancilla's for mixed states) whereas highly non-uniform states need to be represented by the sum of many Clifford circuits.   Because of additivity, we expect that the SRE will mainly grow linearly with system size especially for short-ranged correlated states such as matrix product states (MPS)~\cite{chen2024magic}.

It is therefore important to be able to compute properties of quantum magic such as the SRE.  While recent work has demonstrated an effective approach which scales well for the ground states of one-dimensional (1D) systems, the applications to 2D ground states or finite temperature states of any dimensions are far less efficient.  These works require the construction of tensor network representation of the states~\cite{Oliviero2022,Haug2023,Lami2023,Tarabunga2023,tarabunga2024nonstabilizerness};  
currently the 2D calculations rely on representing the ground states by tree tensor networks whose entanglement structures are incompatible with 2D systems~\cite{Tarabunga2023}. Moreover,the methods based on explicit sampling of Pauli strings have a sampling complexity which grows exponentially with quantum magic for ($n\neq1$)-th SRE, and direct contraction of the tensor network scales with a high polynomial of the bond-dimension, leaving the computations on states with high magic or entanglement as also inefficient~\cite{Haug2023,Lami2023,Tarabunga2023}.  
While tensor network methods are often a powerful approach for 1D ground states, finite temperature and 2D systems are often more tractable by quantum Monte Carlo (QMC).  Various QMC methods exist for calculating the entanglement entropy~\cite{Hastings2010,Luitz2014,Inglis2013,Humeniuk2012,Zhang2011,song2023resummationbased,demidio2024entanglement,JonathanPRL2024}, which is an exponential observable suffering from exponentially vanishing signal-to-noise ratio (SNR, defined as the square of mean over the variance) in naive sampling/computation, and one needs the incremental approaches to convert its SNR back to that of common observables in QMC with polynomial complexity~\cite{ZhangPan2024,ZhouMeng2024}. One of these approaches works by measuring the non-equilibrium work in the partition function space~\cite{Alba2017,DEmidio2020,Zhao2022,Pan2023} with other techniques working by directly sampling reweighted observables from equilibrium partition functions~\cite{song2023resummationbased,ZhangPan2024,ZhouMeng2024,demidio2024entanglement,JonathanPRL2024}. 
In this work, we propose an algorithm applying the non-equilibrium QMC framework for the calculation of SRE in spin systems, which overcomes the aforementioned limitations from tensor-network based methods and broadens the range of accessible quantum systems significantly.

\section{Methods}
The $n$-th SRE is defined as~\cite{Leone2022}: 
\begin{equation} 
\begin{split}
\widetilde{M}_n(\rho)=&\frac{1}{1-n}\ln\sum_{\sigma\in\mathcal{P}_N}\frac{1}{2^N}\Tr[\rho\sigma]^{2n}-S_n(\rho) \\
\equiv&M_n(\rho)-S_n(\rho)
\label{eq:SRE_def}
\end{split}
\end{equation} 
where $N$ is number of spins, $\mathcal{P}_N=\{I,X,Y,Z\}^{\otimes N}$ is the set of all $N$-qubit Pauli strings, and $S_n(\rho)$ is $n$-th entanglement R\'enyi entropy (ERE), i.e., $S_n(\rho)=\ln\Tr\left(\rho^n\right)/(1-n)$, which vanishes for pure states; we will consider both the finite-temperature ($T>0$) and zero-temperature (ground state, $T=0$) cases where the density matrix $\rho$ is 

\begin{equation}
\begin{split}   
\rho=&\frac{e^{-\beta\hat{H}}}{\Tr e^{-\beta\hat{H}}}, \quad T>0, \\
\rho=&\lim_{m \rightarrow \infty} \frac{(-\hat{H})^m\ket{\phi}\bra{\phi}(-\hat{H})^m}{\bra{\phi}(-\hat{H})^{2m}\ket{\phi}}, \quad T=0.
\end{split}
\label{eq:rho}
\end{equation}
where $\hat{H}$ is the Hamiltonian, $\beta=1/T$, and $\ket{\phi}$ is a specialized trial state for projector QMC. 

Essentially, SRE is the R\'enyi entropy of the square coefficients from the decomposition of a quantum state in terms of Pauli strings quantifying  the non-uniformity of Pauli string coefficients, in spirit similar to the inverse participation ratio used in many-body localization~\cite{wegner1980inverse}. For a mixed state, SRE is zero when its purification is a stabilizer state and therefore can be represented with a stabilizer tableau~\cite{Aaronson2004}.  Instead, in the case when a mixed state can be decomposed as a mixture of pure stabilizer states, SRE can be nonzero because the probability distribution over each stabilizer state contributes to SRE as classical correlation. However, for mixed states, SRE is still informative regarding the classical simulability with stabilizer states, when one cares about how many classical resources using stabilizer tableau are required for an accurate description of quantum states.

\begin{figure}
\centering
\includegraphics[width=0.5\textwidth]{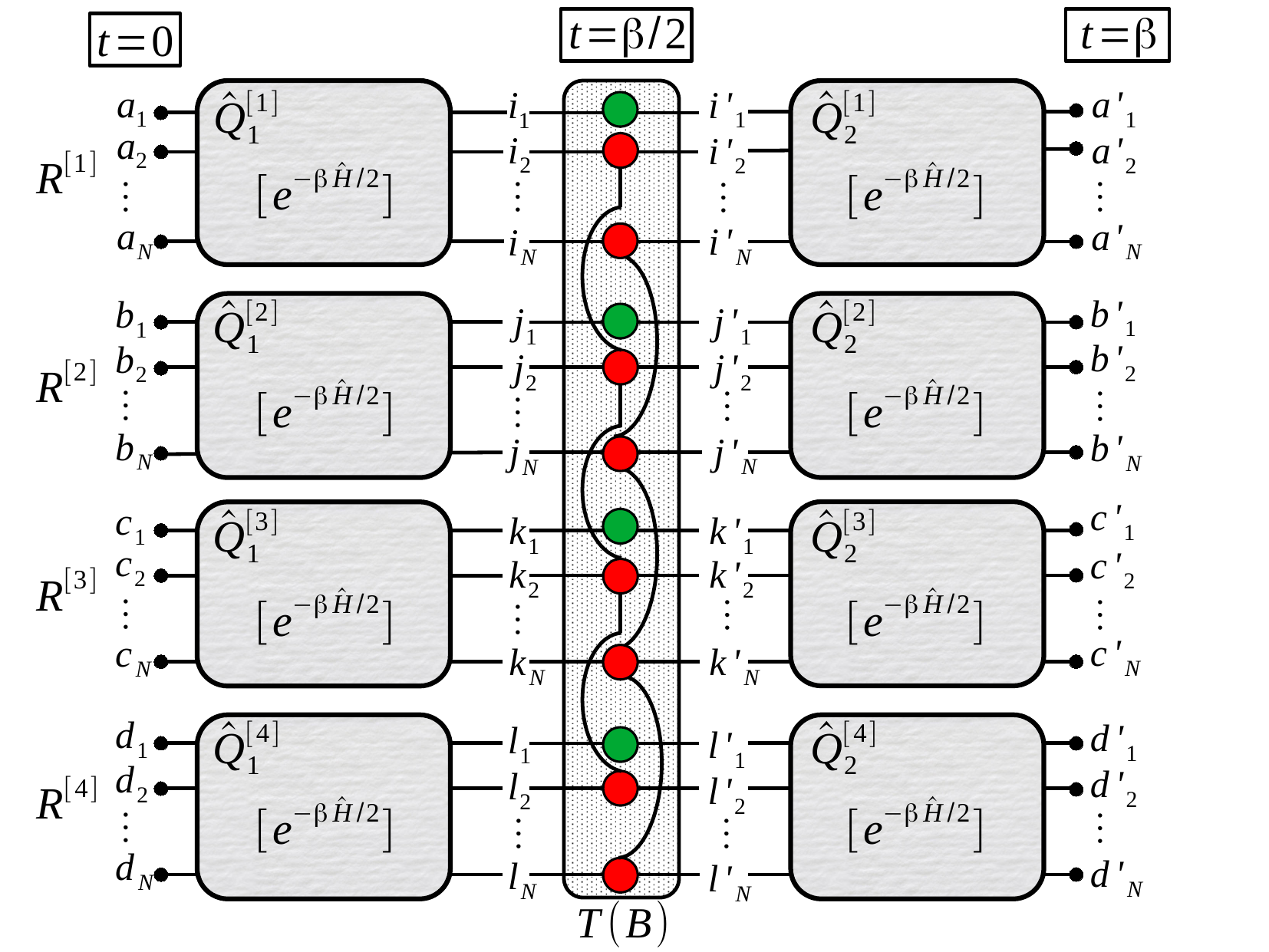}
\caption{Tensor-network diagram for the partition function $\mathcal{Z}_B$ (see Eqn.~\eqref{eq:ZB_def}) with $2n=4$ replicas.  Each (roman indexed) line is a spin in $\{0,1\}$ and gray boxes represent the tensor $e^{-\beta\hat{H}/2}$. $T(B)$ is represented as a $4nN$ tensor consisting of a tensor product of $\mathcal{T}$'s where $B_s=1$,
each of which are represented as connected red dots spanning replicas over their respective site $s$.  All other sites have $B_s=0$ corresponding to identities (green dots).  Note that of the drawn sites, $B_s=1$ on sites 2 and $N$ and $B_s=0$ on site 1.  For finite temperature, the PBC in the imaginary time direction imposes the constraints $\bm{a}=\bm{a}',\bm{b}=\bm{b}',\cdots$, while for the zero temperature case, the sites $\bm{a},\bm{b},...,\bm{a}',\bm{b}',...$ are contracted with $\ket{\phi}$, e.g., $\braket{\bm{a}}{\phi}$.  When this tensor network is sampled with QMC, all spins are sampled and the operator list 
$\hat{Q}=\{\hat{Q}^{[r]}_{1,2},r=1,2,\cdots,2n\}$ is sampled from each gray box. }
\label{fig:keyPic}
\end{figure}

While the previous sampling-based tensor-network methods evaluate Eqn.~\eqref{eq:SRE_def} by directly sampling the Pauli strings, which is responsible for the exponential sample complexity w.r.t. system size $N$, the algorithm presented in this work contracts the summation over the Pauli strings first, and expresses Eqn.~\eqref{eq:SRE_def} in terms of free energy difference between two partition functions instead, and finally adapts the non-equilibrium QMC framework to produce a precise evaluation, which has a polynomial sample complexity. 


The first term in Eqn.~\eqref{eq:SRE_def} for $T>0$ states can be written (combining Eqn.~\eqref{eq:SRE_def} and \eqref{eq:rho}) in terms of the ratio of two partition functions:
\begin{equation}
\begin{split}   
M_n(\rho)
=&\frac{1}{1-n}\ln\frac{\sum_{\sigma\in\mathcal{P}_N}\frac{1}{2^N}\Tr[e^{-\beta\hat{H}}\sigma]^{2n}}{\left(\Tr e^{-\beta\hat{H}}\right)^{2n}}\\\equiv&\frac{1}{1-n}\ln\frac{\mathcal{Z}_{[N]}}{\mathcal{Z}_{\varnothing}}
\label{eq:sre_new}
\end{split}
\end{equation}
Writing
\begin{equation}
\begin{split}    
\mathcal{Z}_{[N]}
&=\Tr\left[\left(e^{-\beta\hat{H}/2}\right)^{\otimes 2n}\left(\frac{1}{2^N}\sum_{\sigma\in\mathcal{P}_N}\sigma^{\otimes 2n}\right)\left(e^{-\beta\hat{H}/2}\right)^{\otimes 2n}\right]
\label{eq:Z_1over2n}
\end{split}
\end{equation}
we can then rewrite the summation over exponentially many Pauli strings as 
\begin{equation}
\begin{split}    
\sum_{\sigma\in\mathcal{P}_N}\sigma^{\otimes 2n}&=\otimes_{s=1}^{N}\left(\sum_{\sigma_s\in\{I,Z,X,Y\}} \sigma_s^{\otimes 2n} \right)\\
&\equiv \otimes_{s=1}^{N} \mathcal{T}
\end{split}
\end{equation}
where the connection tensor $\mathcal{T}$ is a $4n$-leg tensor connecting spins of the same site from each of the $2n$ replicas, whose tensor elements are  
\begin{equation}
\mathcal{T}^{j_1,j_2,\cdots,j_{2n}}_{k_1,k_2,\cdots,k_{2n}}=\prod_{r=1}^{2n}I_{j_r,k_r}+\prod_{r=1}^{2n}Z_{j_r,k_r}+\prod_{r=1}^{2n}X_{j_r,k_r}+\prod_{r=1}^{2n}Y_{j_r,k_r}    
\label{eq:connect_tensor}
\end{equation}
and which can be represented with a tensor-network diagram as
\begin{equation}
\includegraphics[width=0.45\textwidth, valign=c]{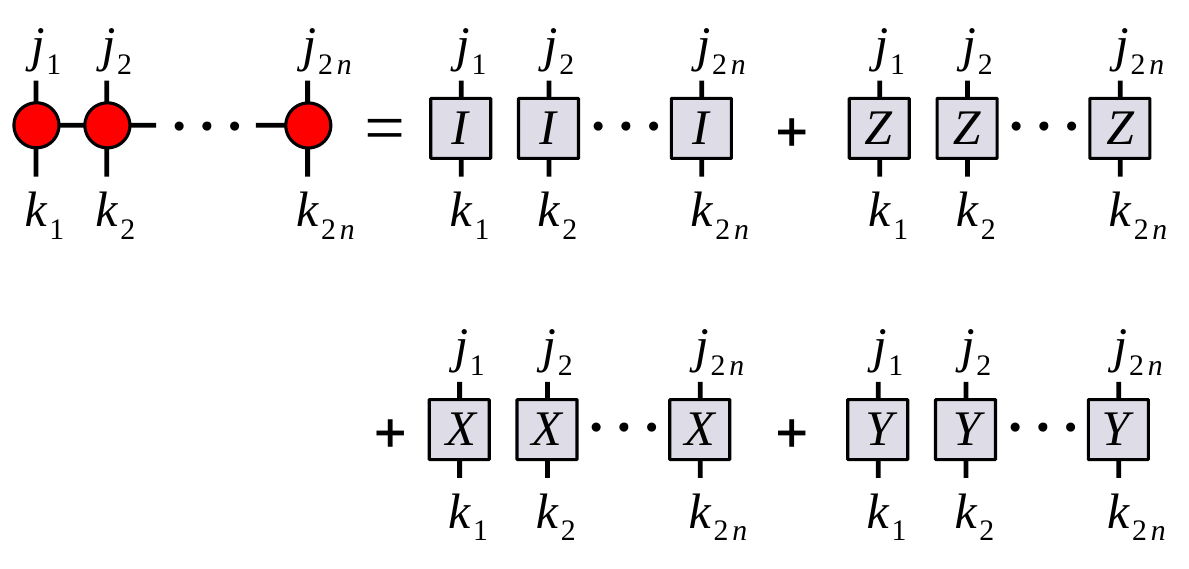}
\label{eq:Diagram_connect_tensor}
\end{equation}
Using the notation that $\mathcal{T}^0=I^{\otimes 2n}$ is the identity for each of the $2n$ replicas, we can define the tensor 
\begin{equation}
T(B)=\frac{1}{2^{N_B}}\otimes_{s=1}^N \mathcal{T}^{B_s}   
\label{eq:def_TB}
\end{equation}
where $B$ is a set of sites (to be thought of as the sites where are connected over the replicas by $\mathcal{T}$) and
$B_s=1$ if site $s\in B$ and $0$ otherwise, $N_B=|B|$ is the total number of sites in $B$.  This lets us write 
\begin{equation}
\mathcal{Z}_B=\Tr\left[\left(e^{-\beta\hat{H}/2}\right)^{\otimes 2n}T(B)\left(e^{-\beta\hat{H}/2}\right)^{\otimes 2n}\right]    
\label{eq:ZB_def}
\end{equation}
and define both $\mathcal{Z}_{[N]}$ and $\mathcal{Z}_{\varnothing}$ where $[N]$ is the set of all sites and $\varnothing$ is the empty set
$\mathcal{Z}_{B}$ for sets of intermediate size will be useful in developing an algorithm to compute the ratio $\mathcal{Z}_{\varnothing}/\mathcal{Z}_{[N]}$.  

Although we have expressed the partition function with a concise tensor-network representation, as shown in Fig.~\ref{fig:keyPic} for $\mathcal{Z}_B$, such a tensor network cannot be directly contracted in an efficient way for large system size due to the presence of high-rank tensor $e^{-\beta\hat{H}/2}$.  

However, one can always view this tensor network contraction as a sum over terms which can then be sampled by Monte Carlo.  
A naive approach to this sampling is to sample terms from $\mathcal{Z}_{N}$  and then count how many of these terms appear in $\mathcal{Z}_{\varnothing}$~\cite{Hastings2010,Zhang2011}.  This can be done in the standard QMC fashion where one samples operator lists for $e^{-\beta\hat{H}/2}$~\cite{Sandvik2010,Sandvik2003,Melko2005}. 
Unfortunately, for the natural case where the SRE grows linearly with system size,  the ratio $\mathcal{Z}_{\varnothing}/\mathcal{Z}_{[N]}$ will get exponentially small, and one will only exponentially rarely see a term which simultaneously belongs to $\mathcal{Z}_{[N]}$ and $\mathcal{Z}_{\varnothing}$  resulting in an exponential  computational cost~\cite{Hastings2010,Zhang2011}. This is the exponential observable problem encountered in the ERE computation and it has been solved recently by sampling reweighted observable with polynomial complexity~\cite{DEmidio2020,Zhao2022,Pan2023,song2023resummationbased,ZhangPan2024,ZhouMeng2024,demidio2024entanglement,JonathanPRL2024}.

We apply the non-equilibrium QMC algorithm to evaluate the ratio $\mathcal{Z}_{\varnothing}/\mathcal{Z}_{[N]}$ efficiently, which amounts to sampling a set of  paths which evolve from $\mathcal{Z}_{\varnothing}$ and $\mathcal{Z}_{[N]}$. To this end, we introduce a parametrized partition function interpolating these two partition functions through a weight function $g(\lambda,N_B)\equiv\lambda^{N_B}(1-\lambda)^{N-N_B}$~\cite{DEmidio2020}:
\begin{equation}
\mathcal{Z}(\lambda)=\sum_{B\subseteq[N]}g(\lambda,N_B)\mathcal{Z}_B,\quad \lambda\in[0,1]   
\label{eq:Zlambda}
\end{equation}

Henceforth, we can rewrite Eq.~\eqref{eq:sre_new} in an integral form:
\begin{equation}
M_n(\rho)=\frac{1}{1-n}\int_0^1 d\lambda\frac{\partial\ln\mathcal{Z}(\lambda)}{\partial\lambda}=-\frac{\Delta F}{1-n}
\label{eq:Mn_integral_0}
\end{equation}
where $\Delta F \equiv F(1) - F(0)$ and $F(\lambda)=-\ln \mathcal{Z}(\lambda).$   The evaluation of the free energy difference can be computed with Jarzynski’s equality~\cite{Jarzynski1997} as $\ln \langle e^{-W} \rangle$ where the work $W$ is the path integral of the derivative of the free energy with respect to the  external parameter $\lambda$, 
\begin{equation}
\begin{split}    
W=&\int_{t_i}^{t_f}dt\frac{d\lambda}{dt}\frac{\partial F(\lambda(t))}{\partial\lambda}\\
=&-\int_{t_i}^{t_f}dt\frac{d\lambda}{dt}\frac{\partial\ln g(\lambda(t),N_B(t))}{\partial\lambda}
\end{split}
\label{eq:work}
\end{equation}
and where $\langle ... \rangle$ denotes the average over all paths which evolve configurations from $\mathcal{Z}(0)$ to $\mathcal{Z}(1)$.

Our final step is thus to use QMC to sample these paths and evaluate the work $W$ from each path. This is done through random walkers where we slowly increase $\lambda$ from 0 to 1, and record the work $W$ as the increment of $\Delta\ln g$. Each random walker samples configurations from an extended configuration space $\varkappa\equiv\{\bm{a},\bm{b},\cdots \bm{a'},\bm{b'},\hat{Q},B\}$ which include the legs of the tensors $\exp(-\beta \hat{H}/2)$ (lowercase letters $\bm{a},\bm{b},\cdots \bm{a'},\bm{b'}$), the internal operator lists ($\hat{Q})$ needed to sample from each replica of $\exp(-\beta \hat{H}/2)$ and the set of connected sites ($B$) (see Fig.~\ref{fig:keyPic}).  The configurations are sampled with probability 
\begin{equation}
\begin{split}    
\text{Wt}(\varkappa=\{\bm{a},\bm{b},\cdots,\hat{Q},B\};\lambda)=&g(\lambda,N_B)\expect{\bm{a}}{\hat{Q}^{[1]}_1}{\bm{i}}\expect{\bm{b}}{\hat{Q}^{[2]}_1}{\bm{j}}\cdots \\
&T(B)^{\bm{i},\bm{j},\cdots}_{\bm{i}',\bm{j}',\cdots}\\
&\expect{\bm{i}'}{\hat{Q}^{[1]}_2}{\bm{a}'}\expect{\bm{j}'}{\hat{Q}^{[2]}_2}{\bm{b}'}\cdots
\end{split} 
\label{eq:weight}
\end{equation}
where the tensor elements of $T(B)$ are
\begin{equation}
T(B)^{\bm{i},\bm{j},\cdots}_{\bm{i}',\bm{j}',\cdots}=\frac{1}{2^{N_B}}\prod_{s\in B}\mathcal{T}^{i_s,j_s,\cdots}_{i'_s,j'_s,\cdots};   
\end{equation}
note that these tensor elements are either 1 if the spin states $\bm{i},\bm{j},\cdots;\bm{i}',\bm{j}',\cdots$ correspond to non-zero elements for each of $\mathcal{T}$, or 0 otherwise. When $n=2$, for example, only 16 out of the 256 elements of $\mathcal{T}$ are non-zero (and equal to 2), such that the sampling space is well constrained and the sign problem will not be induced by the use of $\mathcal{T}$. To get the samples from $\lambda=0$ which are the initial configurations of each random walker, we can use standard QMC on each replica $\exp(-\beta \hat{H}/2)$. 

The basic implementation of the non-equilibrium QMC algorithm for finite-$T$ systems is as follows:
\begin{enumerate}
\item\label{step1} Initial thermalization: perform QMC simulation on $\Tr e^{-\beta\hat{H}}$;
\item\label{step2} Main walker thermalization: collect $2n$ configurations from Step~\eqref{step1} to form the initial configuration for $\mathcal{Z}(\lambda(t_i))$, update the configuration by QMC simulation on $\mathcal{Z}(\lambda(t_i))$;
\item\label{step3} Side walker non-equilibrium process: Collect one configuration from Step~\eqref{step2} as the start;
\begin{enumerate}
\item\label{step3a} Connection topology update: sweep through all the lattice sites, choose to connect ($P_{\text{c}}$) or disconnect ($P_{\text{d}}$) each spin from the replicas based on the probability:
\begin{equation}
P_{\text{c}}=\min\left\{\frac{\lambda}{1-\lambda},1\right\},\quad P_{\text{d}}=\min\left\{\frac{1-\lambda}{\lambda},1\right\}    
\end{equation}
if the spin states from all the replicas have correspond to the non-zero entries of the connection tensor $\mathcal{T}$.
\item Configuration update: perform QMC update on the connected replicas;
\item\label{step3c} $\lambda$ increment and observable: increase $\lambda$ by a small amount $\lambda_{i+1}\leftarrow \lambda_{i}+\Delta\lambda$. Record the change of $\ln g(\lambda,N_B)$ at this $i$-th step: $\Delta\ln g(\lambda_i)=
\ln g(\lambda_{i+1},N_B(\lambda_{i}))-\ln g(\lambda_{i},N_B(\lambda_i))$.
\end{enumerate}
Repeat the Step~\eqref{step3a}-\eqref{step3c} until $\lambda=\lambda(t_f)$. The overall work done in this path is given by $W=-\sum_i\Delta\ln g(\lambda_i)$. 
\item\label{step4} Repeat Step~\eqref{step3} to collect a sample set of $W$. Estimate SRE using Eqn.~\eqref{eq:Mn_integral_0}.
\end{enumerate}

Moreover, we note two parallelization strategies to speed up this algorithm: (1) Divide the evolution of $\lambda$ into $K$ smaller intervals: $[0,1]=[0,1/K]\cup[1/K,2/K]\cup\cdots\cup[1-1/K,1]$; Apply the above algorithm to each interval independently (e.g., $t_i=k/K$, $t_f=(k+1)/K$); Then the original SRE is given by the sum from these smaller intervals~\cite{Zhao2022}. (2) As each side walker from Step~\eqref{step3} evolves independently, one can distribute each of them to different processors and only collect the observable $W$ from them at the end of the evolution. 

This new framework we have developed which represents the ratio of partition functions as tensor networks and then evaluates that ratio using non-equilibrium QMC can be generalized to various other quantities including other types of R\'enyi entropy calculations, 
e.g., entanglement R\'enyi negativity (ERN), a measure of quantum correlation within mixed states~\cite{Wu2020,Lu2020,Calabrese2012,calabrese2013entanglement,alba2013entanglement,Chung2014}, and Shannon R\'enyi entropy, also known as participation R\'enyi entropy (PRE), a measure for the wave function localization in the computational basis~\cite{Stephan2009,Stephan2010,Zaletel2011,Stephan2014}. We refer to the Appendix.~\ref{append:framework} for discussion of these generalizations.


\section{Numerical Experiments}
We benchmark the algorithm for 2nd SRE ($n=2$) of the transverse field Ising model, i.e.,
\begin{equation}
\hat{H}=-J\sum_{\langle i,j\rangle}Z_iZ_j-h\sum_i X_i
\end{equation}
where stochastic series expansion (SSE) is applied to update the configurations~\cite{Sandvik2010,Sandvik2003,Melko2005,Wang1987}. 

\subsection{Implementation of SSE on Ising Model - Finite Temperature}
{\label{append:sseT}}
In this section, we describe how SSE works for the non-equilibrium algorithm on the transverse field Ising Model ~\cite{Sandvik2010,Sandvik2003,Melko2005,Wang1987}. 

The configuration space for SSE consists of an initial product state for the $N$ spins and an operator list selected from 
\begin{equation}
\begin{split}
H_{0,0}&=1;\quad H_{k,0}=hX_k;\quad H_{k,k}=hI_k; \\ H_{k,j}&=|J|+JZ_kZ_j;\quad k,j>0 \text{ and } k\neq j.    
\end{split}
\end{equation}
Because of the presence of the connection tensor $\mathcal{T}$ in our algorithm, the initial product state is the spin configuration at the location of $\mathcal{T}$ - i.e. fixing $\bm{i},\bm{j},\bm{k},\bm{l},\bm{i'},\bm{j'},\bm{k'},\bm{l'}$. 

There are two stages of configuration updates within SSE: diagonal update and off-diagonal update. The diagonal update here is identical to the standard one~\cite{Sandvik2010}. For the off-diagonal update, we use the cluster updating scheme, where we group the legs from all the operators into clusters, each of them containing the whole two-spin operators but ending at single-spin operators. After all clusters are formed, we choose to flip each cluster with $1/2$ probability following the Swendsen-Wang scheme~\cite{Wang1987}. 

\begin{figure}
\centering
\includegraphics[width=0.35\textwidth]{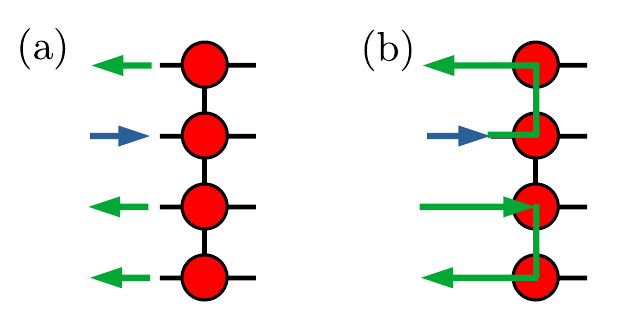}
\caption{Schematic diagrams for (a) cluster updating and (b) loop updating when the cluster or the directed loop reaches the leg of $\mathcal{T}$. } 
\label{fig:sseloops}
\end{figure}

When a cluster reaches one of the $2n$ legs from $\mathcal{T}$, we use the multi-branch clustering to continue growing the cluster out of $\mathcal{T}$~\cite{Melko2005}. More specifically, the cluster will branch out from the $n-1$ legs of $\mathcal{T}$ on the same side as the reached leg, so that the cluster enters into other replicas. This ensures that after the update, the spin configuration still corresponds to a non-zero entry of $\mathcal{T}$. For example, as shown in Fig.~\pref{fig:sseloops}{a}, when a cluster from the second replica reaches one leg of $\mathcal{T}$ (indicated by the blue arrow going towards $\mathcal{T}$), the cluster grows into the rest of the replicas (as indicated by the green arrows going outwards from $\mathcal{T}$). Similarly, in Fig.~\pref{fig:sseloops}{b}, we also show the diagram for the directed loop update scheme if we apply the algorithm to models such as XXZ model, where we let the directed loop from the second replica grow into the first replica, and a new directed loop form within third and forth replicas.

\subsection{Implementation of SSE on Ising Model - Zero Temperature}
{\label{append:SSET0}}
To obtain the zero-temperature or the ground state SRE on the Ising model, we use projector QMC with the product state $\ket{00\cdots0}$ as the initial trial state $\ket{\phi}$ and an operator list with long enough length $m$.  Different from the finite-$T$ algorithm, here there are open boundary conditions on the imaginary time axis,  that is, the cluster growth also ends at the boundary. 
In the zero temperature situation, we fix the initial product state of $|00...0\rangle$ at $\bm{a},\bm{b},\bm{c},\bm{d}$ and $\bm{a'},\bm{b'},\bm{c'},\bm{d'}$. 
Note that this gives the SRE of the ground state which has overlap with the $|00...0\rangle $ sector.

\begin{figure}
\centering
\includegraphics[width=0.5\textwidth]{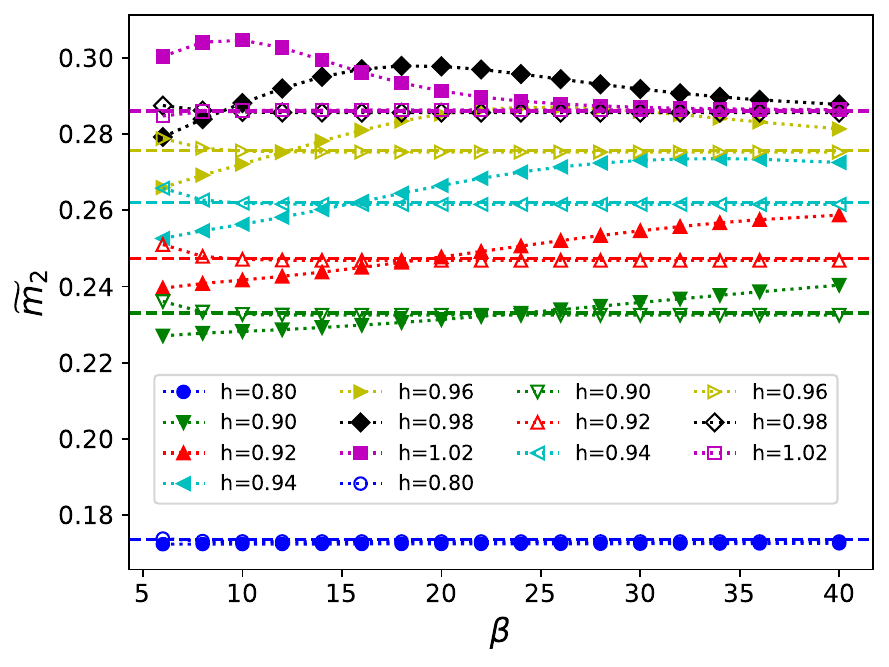}
\caption{Ground state 2nd SRE density on 1D $N=20$ Ising model (PBC, $J=1$). Here the data are obtained from exact contractions of MPS, see Suppl. Sec.V. The ways to obtain MPS are different: the horizontal dashed lines come from DMRG; circles come from imaginary time evolution starting from $\ket{00\cdots0}$; triangles come from imaginary time evolution starting from $\frac{1}{2^{N/2}}\sum_{z\in\{0,1\}^N}\ket{z}$.  } 
\label{fig:sre_iT_init}
\end{figure}

We could also choose the equal superposition of all product states, i.e., $\ket{\phi}=\frac{1}{2^{N/2}}\sum_{z\in\{0,1\}^N}\ket{z}$ as the trial state, which in fact converges to the ground state SRE value faster, as indicated from Fig.~\ref{fig:sre_iT_init} for the case of imaginary-time evolved MPS. However, within SSE, it will not directly give the SRE for pure states.

To illustrate this point, let us consider a simple example. At $h=0$, after initial thermalization, the initial product states are $\ket{00\cdots0}$ or $\ket{11\cdots1}$ with equal probabilities, and the operators that flip the spins ($H_{k,0}$) will not show up in the operator lists. Consequently, SSE only gives the configurations corresponding to the mixed states $\rho=(\ket{00\cdots0}\bra{00\cdots0}+\ket{11\cdots1}\bra{11\cdots1})/2$, which gives $\widetilde{M}_2(\rho)=\ln 2$, while either $\ket{00\cdots0}$ or $\ket{11\cdots1}$ gives $\widetilde{M}_2(\rho)=0$ as pure states. Because of the ground degeneracy, for a pure state $\ket{g}=\alpha\ket{00\cdots0}+\beta\ket{11\cdots1}$, SRE varies with the choice of coefficients $\alpha$ and $\beta$. To resolve these issues, we fix the initial product state as $\ket{00\cdots0}$, giving an unambiguous evaluation of SRE. Note that the above discussions also applies to the case of $h\ll1$.

\subsection{Numerical Results}

\begin{figure*}
\centering
\includegraphics[width=\textwidth]{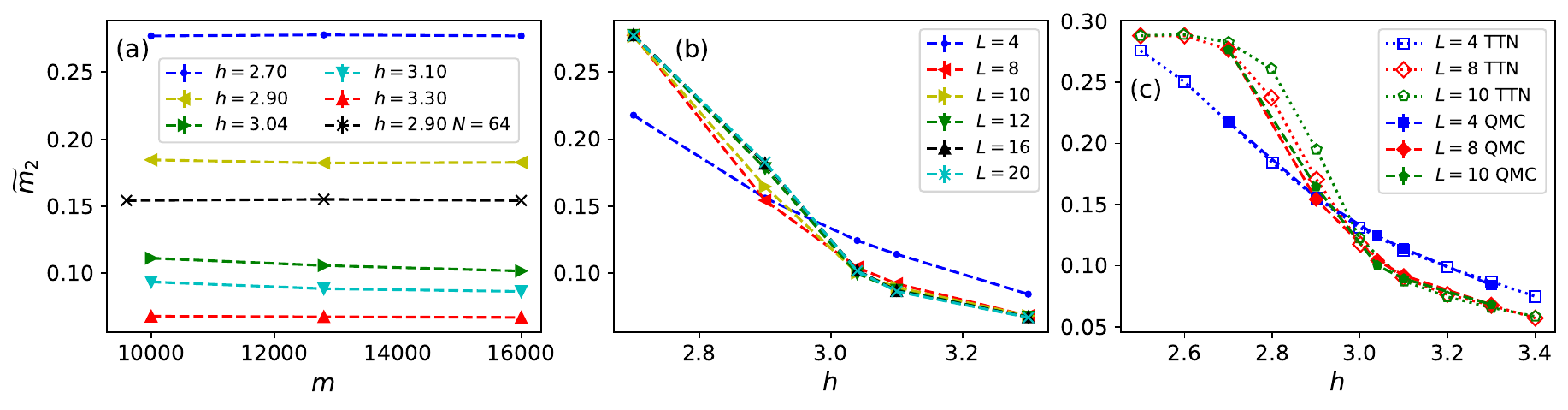}
\caption{(a) Convergence of 2nd SRE density on ground state of 2D transverse field Ising model (PBC, $J=1$, $N=400$ with dot markers, $N=64$ with cross markers), w.r.t. the length of operator list $m$. (b) 2nd SRE density on ground state of 2D transverse field Ising model (PBC, $J=1$) versus the strength of field $h$. (c) Comparison between the QMC data (dot markers) and the data from Ref.~\cite{Tarabunga2023} using TTN (square markers).}
\label{fig:sre_2d_compare}
\end{figure*}

For ground state SRE, one needs to use sufficiently large length of operator list $m$ to ensure the convergence towards the exact ground state SRE. In Fig.~\pref{fig:sre_2d_compare}{a}, we show the SRE density at the chosen $m\in\{9600, 10000, 12800, 16000\}$, where we observe that the SRE are well converged, despite that for some of them (e.g., $N=400$ at $h=3.10$ and $h=3.30$), the SRE decrease slightly from $m=10000$ to $m=16000$.
In Fig.~\pref{fig:sre_2d_compare}{b}, we show the SRE density $\widetilde{m}_2\equiv\widetilde{M}_2/N$ versus $h$, and observe that SRE densities intersect at different $h$ at larger system size from that at smaller system size. In Fig.~\pref{fig:sre_2d_compare}{c}, we directly compare our results with those from Ref.~\cite{Tarabunga2023} using tree tensor networks (TTN), which are in good agreement except for at $h=2.9$ at large $L$. 

For the rest of the numerical experiments, we choose $m$ between $10000$ and $16000$, and use 160 path samples (i.e. the number of non-equilibrium work iterations) for ground state calculations and 640 for finite temperature calculations for all the system sizes considered here. We choose $\Delta\lambda=10^{-4}$. For a single path-sample on our largest system of $N=16 \times 16$ computing the ground state SRE takes  $\lesssim4$ CPU hours using a code implemented in C++. 

\begin{figure*}
\centering
\includegraphics[width=\textwidth]{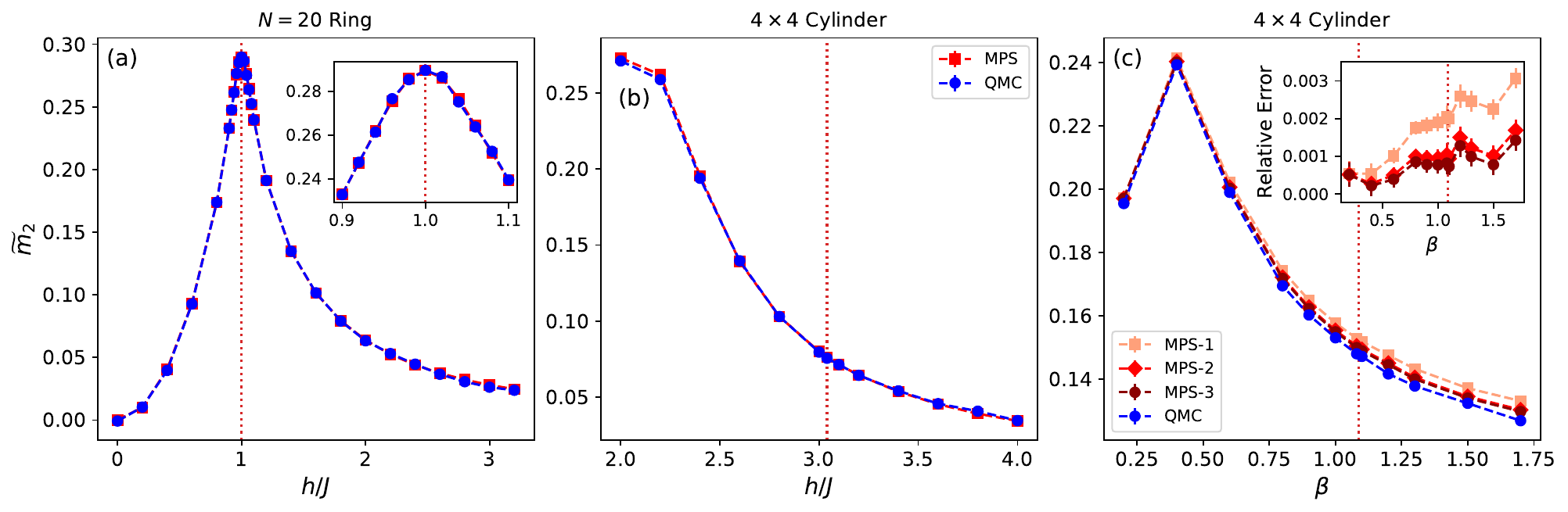}
\caption{Comparisons of 2nd SRE density on the transverse-field Ising model between our algorithm and MPS-based algorithms. (a) Ground state SRE of 1D Ising model (PBC) with 20 spins. Inset: the zoom-in near the phase transition. (b) Ground state SRE of $4\times4$ 2D Ising model (PBC-OBC). (c) Finite-$T$ SRE of $4\times4$ 2D Ising model (PBC-OBC) at $h/J=2.75$, for which the phase transition happens at $\beta=1.0874(1)$\cite{Wu2020,Hesselmann2016}. The MPS data are produced with different bond dimensions $\chi$ and Trotter time intervals $\tau$: MPS-1 (light red): $\chi=256$, $\tau=5\cdot 10^{-3}$; MPS-2 (red): $\chi=512$, $\tau=2.5\cdot 10^{-3}$; MPS-3 (dark red): $\chi=640$, $\tau=2\cdot 10^{-3}$. Inset: the relative error from MPS results to the QMC results. The red vertical dotted lines in (a)-(c) mark the parameter value for phase transitions. Error bars are too small to be visible.}
\label{fig:sre_MPS}
\end{figure*}

In Fig.~\ref{fig:sre_MPS}, we show the comparisons between the 2nd SRE density $\widetilde{m}_2$ from the non-equilibrium QMC and those from matrix product states (MPS) in various scenarios: (a) ground state SRE of 1D Ising ring (i.e., periodic boundary conditions, PBC); (b) ground state SRE of 2D Ising cylinder (i.e., PBC on one spatial direction and open boundary condition on the other, PBC-OBC); (c) finite-$T$ SRE of 2D Ising cylinder. They all display quantitative agreements when comparing against with the results from the tensor-network methods. Remarkably, in case (c), the noticeable discrepancies between the QMC and MPS data result from the errors of limited bond dimension $\chi$ and the finite Trotter splitting time interval $\tau$ for the MPS method as can be seen from the observation that the MPS data converges towards the QMC data as we increase $\chi$ and decrease $\tau$.  This suggests that to reach the same accuracy as QMC, the MPS-based methods requires more computational resources and careful extrapolation. 
It is interesting to observe from Fig.~\ref{fig:sre_MPS} that SRE density peaks at the phase transition for the ground states of the 1D transverse-field Ising model, but not in 2D cases. 
It is an interesting open question to understand this observation.

\begin{figure}
\centering
\includegraphics[width=0.5\textwidth]{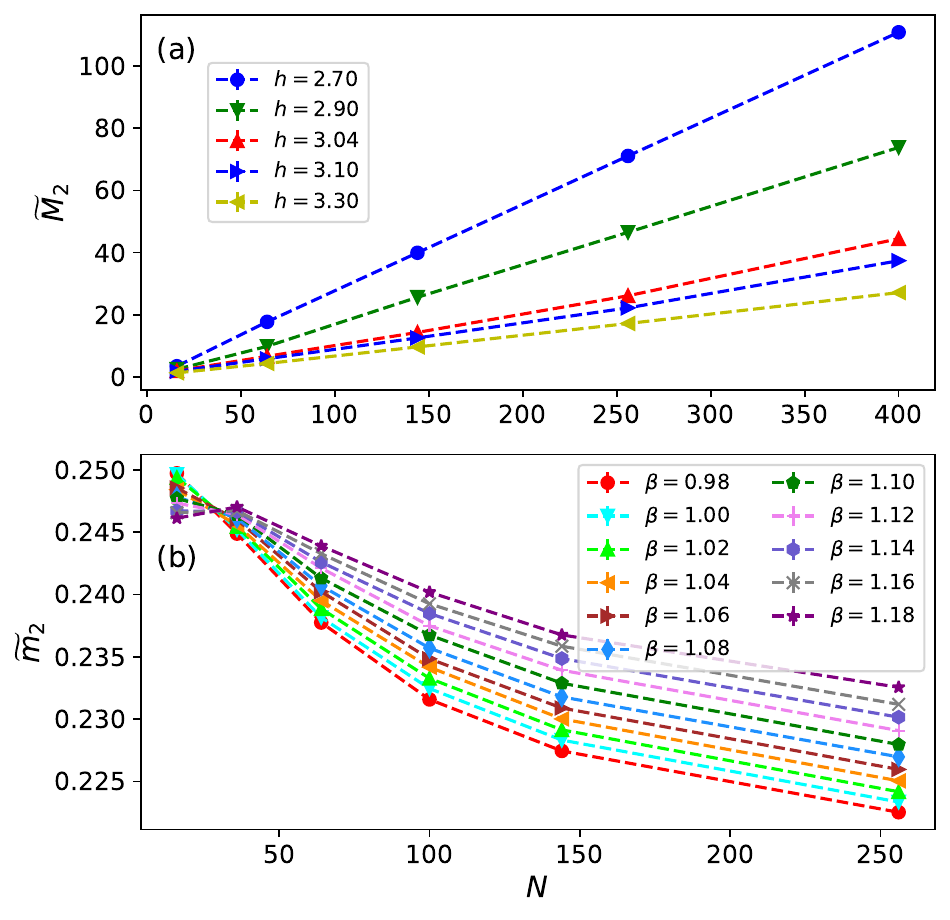}
\caption{2nd SRE on 2D $N=L^2$ transverse-field Ising model (PBC, $J=1$). (a) Ground state SRE; (b) Finite-$T$ SRE density at $h=2.75$.}
\label{fig:sre_largeN}
\end{figure}

\begin{figure}
\centering
\includegraphics[width=0.5\textwidth]{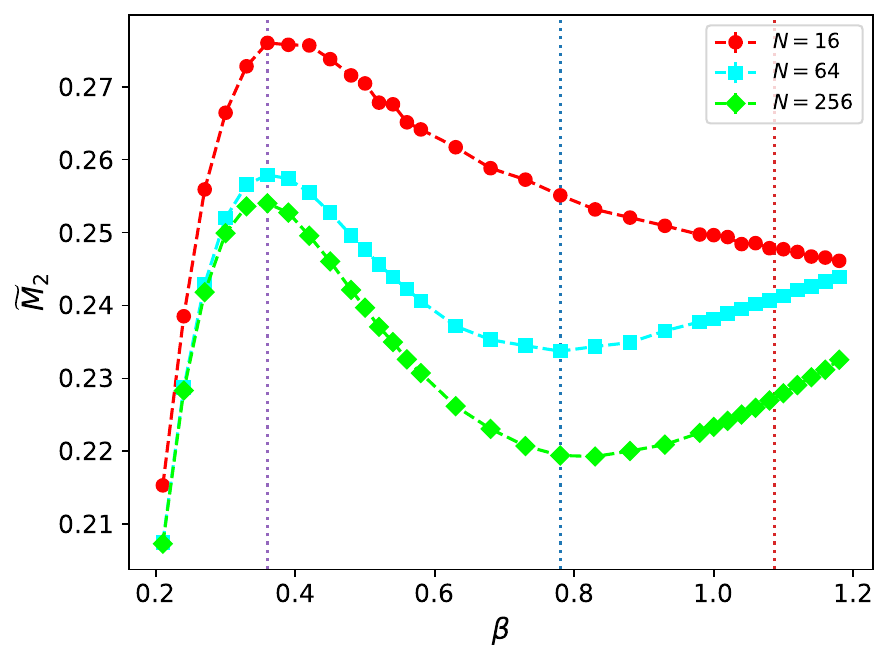}
\caption{2nd SRE on 2D $N=L^2$ transverse-field Ising model (PBC, $J=1$, $h=2.75$) versus $\beta$ The vertical dotted lines mark the extremals of the curves: $\beta_1\simeq0.36$ (purple), $\beta_2\simeq0.78$ (blue) and paramagnetic-ferromagnetic phase transition: $\beta_c\simeq1.0874(1)$ (red).}
\label{fig:sre_morebeta}
\end{figure}

In Fig.~\ref{fig:sre_largeN}, we report our results on larger systems, and we are able to draw the following conclusions: (1) 2nd SRE mainly grows linearly with system size for short-range correlated pure states, as indicated by the ground states on 2D transverse-field Ising model. (2) At $T>0$, there exist subleading terms from 2nd SRE weakly dependent on $T$, as evident from the crossing in Fig.~\pref{fig:sre_largeN}{b}. 
In Fig.~\ref{fig:sre_morebeta}, we show the 2nd SRE versus a broader range of temperature, and find two extremals at $\beta\simeq0.36$ and $0.78$ for sufficiently large systems.
It is unclear whether the $\beta$ of these extremals have any particular physical significance.


\section{Computational Cost}
In this section, we analyze the computation cost of the proposed QMC algorithm, defined as the product of the sample complexity and the time complexity for each sample. 

For finite-$T$ systems, the time complexity to evaluate a single observable $W$ comes from doing a single update of finite temperature SSE on $2nN$ spin particles at all $1/\Delta \lambda$ number of $\lambda$'s.  Presumably $N\Delta\lambda\sim\mathcal{O}(1)$, however, we choose a fixed large enough  $1/\Delta\lambda$  for all the system sizes in our implementations.  In SSE, the time $\mathcal{C}_{\mathcal{T}}$ and memory $\mathcal{C}_{\mathcal{M}}$ complexity are both linear w.r.t. the length of the operator list, such that we have $\mathcal{C}_{\mathcal{T}}(N,\beta)=\mathcal{O}(\beta N)$, $\mathcal{C}_{\mathcal{M}}(N,\beta)=\mathcal{O}(\beta N)$~\cite{Sandvik2003,Sandvik2010}. 

For $T=0$ or ground state SRE, we use projector QMC and the key difference for the time complexity is that the length of the operator list $m$ has to be sufficiently large to decrease the contributions from the excited states and guarantee the convergence to the ground state.  We argue that the lower bound for $m$ such that the relative bias error to the exact ground state SRE $\widetilde{M}_n(g)=M_n(g)$ is below $\delta_r$ is given by (see SM for the derivation)
\begin{equation}
m\gtrsim \frac{|E_g|}{2\Delta_g}\ln\left(\frac{2nr_0^2}{(n-1)\delta_r\cdot M_n(g)}\right)    
\label{eq:sampleComplex}
\end{equation}
where $E_g$ is the ground state energy, $\Delta_g$ is the energy gap, and $r_0=\braket{e}{\phi}/\braket{g}{\phi}$ where $|g\rangle$ and $|e\rangle$ are the ground and first excited states respectively and   $\ket{\phi}$ is the initial trial state.  Notice that the time complexity here primarily depends inversely with the gap which is well controlled even for gapless second-order transitions where the gap generically decreases polynomially with system size.

For the sample complexity, we empirically observe that the SNR of the samples for the evaluation of $e^{-M_2(\rho)}$ in the 2D Ising model scales as $N^{-\alpha}$ with $\alpha\simeq 1.20$, suggesting the polynomial sample complexity $\mathcal{O}(N^{\alpha})$ regardless of using either finite-$T$ or projector QMC algorithms. Similar behavior of the SNR in the ERE computation has also been discussed and achieved in Refs.~\cite{ZhangPan2024,ZhouMeng2024} (see SM for more details).

\section{Conclusion}
In this work, we develop a novel algorithm to compute the SRE of finite temperature and ground state systems in all dimensions for sign-problem free Hamiltonians.   To accomplish this, we show how to contract a tensor network representation of the SRE  using non-equilibrium QMC. 
It has proven to be accurate and efficient, regardless of the spatial dimensions and temperature of the model, as observed from the numerical experiments on the transverse-field Ising model and the complexity analysis of this algorithm. We expect this algorithm would serve as a useful numerical tool to understand quantum systems from the new perspective of quantum magic. Interesting future directions include, but are not limited to, an in-depth study of phase transition features from $n(\geq2)$th SRE consisting of the derivatives~\cite{Wu2020}, the application to models with deconfined quantum critical points~\cite{senthil2004deconfined,Levin2004,SenthilBalents2004,Wang2017,Qin2017} or field theory~\cite{Fliss2021,bulgarelli2023entanglement,bulgarelli2024duality}, the extension of this approach to Hamiltonians with sign problems, and the exploration of replacing the non-equilibrium QMC algorithm with other efficient algorithms for evaluating the exponential observable~\cite{ZhangPan2024,ZhouMeng2024,demidio2024entanglement}.

\begin{acknowledgments}    
We would like to thank T.C. Lu, Z.Y. Meng, J. Zhao, J. D'Emidio, M. Dalmonte, E. Tirrito, M. Frau and P. Tarabunga for insightful discussions and valuable comments. This work made use of the Illinois Campus Cluster, a computing resource that is operated by the Illinois Campus Cluster Program (ICCP) in conjunction with the National Center for Supercomputing Applications (NCSA) and which is supported by funds from the University of Illinois at Urbana-Champaign. We acknowledge support from the NSF Quantum Leap Challenge Institute for Hybrid Quantum Architectures and Networks (NSF Award No. 2016136).
\end{acknowledgments}


\bibliography{main}

\appendix

\section{Connection of Lattice Spins at the End of Non-equilibrium Process}
{\label{append:proofConnect}}
In the main text, we note that at the end of each side walker for the non-equilibrium process, all the lattice sites will be connected, falling into the configuration space of $\mathcal{Z}_1$. Here we provide an argument that we anticipate that this will happen generically. 

At a given site, the spins of the $2n$ replicas are connectable if the spin configuration at that site corresponds to a non-zero entry of $\mathcal{T}$. We set $q$ as the probability for this kind of spin configurations to show up during the QMC update. For convenience, we also set $p=\frac{1}{\lambda}-1$, and note that $p\rightarrow0$ as $\lambda\rightarrow1^-$, and $q$ should stay finite for $\forall \lambda\in[0,1]$.

Considering a single site $k$, there will be the following four events happening when updating the connection topology at $\lambda>1/2$:
\begin{enumerate}
\item With probability $P_C(\lambda)$, site $k$ is initially connected,
\begin{enumerate}
\item site $k$ will stay connected with probability $1-p$;
\item site $k$ will be disconnected with probability $p$.
\end{enumerate}
\item With probability $P_D(\lambda)$, site $k$ is initially disconnected,
\begin{enumerate}
\item site $k$ will be connected with probability $q$;
\item site $k$ will stay disconnected with probability $1-q$.
\end{enumerate}
\end{enumerate}
This gives the transition equation as 
\begin{equation}
\begin{bmatrix}
P_C(\lambda_{i+1}) \\ P_D(\lambda_{i+1})    
\end{bmatrix}=
\begin{bmatrix}
1-p & q \\ p & 1-q    
\end{bmatrix}
\begin{bmatrix}
P_C(\lambda_{i}) \\ P_D(\lambda_{i})    
\end{bmatrix}
\end{equation}
By diagonalizing the transition matrix, which gives eigenvalues $1$ and $1-p-q$, we will have the (dis)connection probability at $\lambda\rightarrow1$ (after many times of multiplication of the transition matrix) approximately as
\begin{equation}
\begin{bmatrix}
P_C(\lambda) \\ P_D(\lambda)    
\end{bmatrix}\approx
\frac{qP_C(\lambda_0)+pP_D(\lambda_0)}{q^2+p^2}
\begin{bmatrix}
q \\ p    
\end{bmatrix}
\end{equation}
where we have assumed that $p$ and $q$ are independent (or at least weakly dependent) of $\lambda$ after a threshold $\lambda_0$. Since $P_C(\lambda)/P_D(\lambda)\approx q/p\rightarrow\infty$ as $\lambda\rightarrow1$, we conclude that site $k$ will be connected at the end of the non-equilibrium process.

In our implementation, we fix the increment of $\lambda$ to be $\Delta\lambda=10^{-4}\sim10^{-3}$. We observe a small number of paths end with one spin site unconnected. In this case, we will abandon this path, as these paths will give ill-defined work involving $\ln0$. It was showed that a protocol with finer $\Delta\lambda$ near the end points $\lambda=0$ or $\lambda=1$ helps reducing the error bars in equilibrium measurements~\cite{demidio2024entanglement}. We expect that such a protocol will help reducing the rate of paths with unconnected sites at the end of evolution. 

\section{Matrix Product State Based Methods}{\label{append:mps}}
In the main text, we compare the results from the QMC algorithm to those from MPS-based algorithms. Here we will briefly explain the MPS-based algorithms constructed for the benchmark of the QMC algorithm. 

\textit{Exact Tensor Network Contraction}\hspace{0.8 mm} For 1D systems, where MPS can approximate the ground state of the Ising model with relatively small bond dimensions, we can perform the exact tensor network contraction to compute SRE~\cite{Haug2023}. Here DMRG is used to produce the ground state MPS, and then $4n$-replica of it are used together with the connection tensor $\mathcal{T}$ to perform the contraction over the spin bonds and virtual bonds site by site. Note that this method has a memory cost $O(\chi^{4n})$, limiting the allowed bond dimension $\chi\leq12$. Henceforth we only use it to generate data for 1D ground states.

\textit{Pauli String Sampling from Matrix Product State}\hspace{0.8 mm} In 2D systems, we need larger bond dimension to represent the states. Here we use DMRG to obtain MPS for the ground state of the 2D Ising model. Then we use this MPS to sample the Pauli strings autoregressively for the evaluation of SRE~\cite{Lami2023}. Note that during sampling, the tensors from unsampled sites need to contract with $\mathcal{T}$ instead of directly contracting the spin indices. 

\textit{Pauli String Sampling from Matrix Product Density Operator}\hspace{0.8 mm} At finite-$T$, we use MPDO to represent the finite-$T$ state, which is obtained with imaginary time evolution on the maximally mixed states (infinite-$T$ state). The Pauli string samples are again sampled autoregressively from the MPDO~\cite{tarabunga2024nonstabilizerness}. Again, during sampling, the tensors from unsampled sites need to contract with $\mathcal{T}$.

\section{Derivation for Eqn.~\eqref{eq:sampleComplex}}
\label{append:derive_m}
In this section, we show the derivation for Eqn.~\eqref{eq:sampleComplex}, which implies that the time complexity for ground state SRE of non-degenerate systems is polynomial in the system size. 

Suppose that the state after projector QMC is given by
\begin{equation}
\ket{\phi}=\frac{1}{\sqrt{c_g^2+c_e^2}}\left(c_g\ket{g}+c_e\ket{e}\right)    
\label{eq:phi_expand}
\end{equation}
where $\ket{g}$ is the ground state, and $\ket{e}$ is the excited state (or the superposition of excited states, as long as $\braket{e}{e}=1$, $\braket{g}{e}=0$). Without loss of generality, assume $\ket{\phi}$ is defined with real numbers. Denote the energies as $E_g=\expect{g}{\hat{H}}{g}$, $E_e=\expect{e}{\hat{H}}{e}$, $\Delta_g=E_e-E_g$. In other words, the state $\ket{\phi}$ has some error away from the ground state, and next we will show how this error affects the calculation of SRE. From the main text, we have
\begin{equation}
\begin{split}    
e^{(1-n)M_n(\ket{\phi}\bra{\phi})}=&\bra{\phi}^{\otimes 2n}\frac{1}{2^N}\sum_\sigma\sigma^{\otimes 2n}\ket{\phi}^{\otimes 2n}\\
=&\bra{\phi}^{\otimes 2n}T([N])\ket{\phi}^{\otimes 2n}   
\label{eq:Mn_ground}
\end{split}
\end{equation}

By replacing $\ket{\phi}^{\otimes 2n}$ with Eqn.~\eqref{eq:phi_expand}, we will obtain a polynomial expansion of Eqn.~\eqref{eq:Mn_ground} in terms of $\bra{g}^{\otimes n_l}\bra{e}^{\otimes (2n-n_l)}T([N])\ket{g}^{\otimes n_r}\ket{e}^{\otimes (2n-n_r)}$. However, we conjecture that all these terms are small compared to $\bra{g}^{\otimes 2n}T([N])\ket{g}^{\otimes 2n}$ and $\bra{e}^{\otimes 2n}T([N])\ket{e}^{\otimes 2n}$. We heuristically examined this conjecture using exact diagonalization on small 1D Ising model. Henceforth, we end up with a simplified version of Eqn.~\eqref{eq:Mn_ground}
\begin{equation}
\begin{split}    
e^{(1-n)M_n(\ket{\phi}\bra{\phi})}=&\frac{c_g^{4n}}{(c_g^2+c_e^2)^{2n}}\bra{g}^{\otimes 2n}T([N])\ket{g}^{\otimes 2n}+\\
&\frac{c_e^{4n}}{(c_g^2+c_e^2)^{2n}}\bra{e}^{\otimes 2n}T([N])\ket{e}^{\otimes 2n}
\label{eq:Mn_SggSee}
\end{split}
\end{equation}
For convenience, let us denote the relevant quantities as $S_{gg}=\bra{g}^{\otimes 2n}T([N])\ket{g}^{\otimes 2n}$, $S_{ee}=\bra{e}^{\otimes 2n}T([N])\ket{e}^{\otimes 2n}$, $r=c_e/c_g$. If $r\ll1$, we can further ignore the second term in Eqn.~\eqref{eq:Mn_SggSee} and obtain the bias error as 
\begin{equation}
\delta = \frac{\ln\left(1-2nr^2\right)}{1-n}    
\label{eq:abs_bias}
\end{equation}
Eqn.~\eqref{eq:abs_bias} indicates that the bias error is always positive for sufficiently small $r$. In other words, the estimate from projector QMC will provide an upper bound for the exact ground state SRE $M_n(g)$; as we increase the precision of projector QMC (e.g., the length of the operator list in SSE), the estimated SRE $M_n(\ket{\phi}\bra{\phi})$ will decrease towards $M_n(g)$, which is consistent with our observation in the numerical experiments (see Fig.\pref{fig:sre_2d_compare}{a}). 

Subsequently we have the relative bias error as 
\begin{equation}
\delta_r\equiv\frac{|M_n(\ket{\phi}\bra{\phi})-M_n(g)|}{|M_n(g)|}=\frac{|\ln(1-2nr^2)|}{|(1-n)M_n(g)|}    
\end{equation}
Note that in projector QMC, we have the approximation for $r$ as
\begin{equation}
r\simeq r_0\left(\frac{E_g+\Delta_g}{E_g}\right)^m    
\end{equation}
such that we get
\begin{equation}
m\gtrsim \frac{\ln\left(\frac{1-e^{(1-n)\delta_r M_n(g)}}{2nr_0^2}\right)}{2\ln\left( 1+\frac{\Delta_g}{E_g} \right)}  
\end{equation}
At last, assuming $\delta_r\cdot M_n(g)\ll 1$ and $\Delta/|E_g|\ll 1$, we approximate the above result as
\begin{equation}
m\gtrsim \frac{|E_g|}{2\Delta_g}\ln\left(\frac{2nr_0^2}{(n-1)\delta_r\cdot M_n(g)}\right)    
\end{equation}

\section{Statistical Properties of the Non-equilibrium QMC Algorithm}
{\label{append:statQMC}}

In the main text, we have seen that the relative error on the SRE scales polynomially with system size. In this section, we will explore this scaling in various ways. 
\begin{figure*}
\centering
\includegraphics[width=1.0\textwidth]{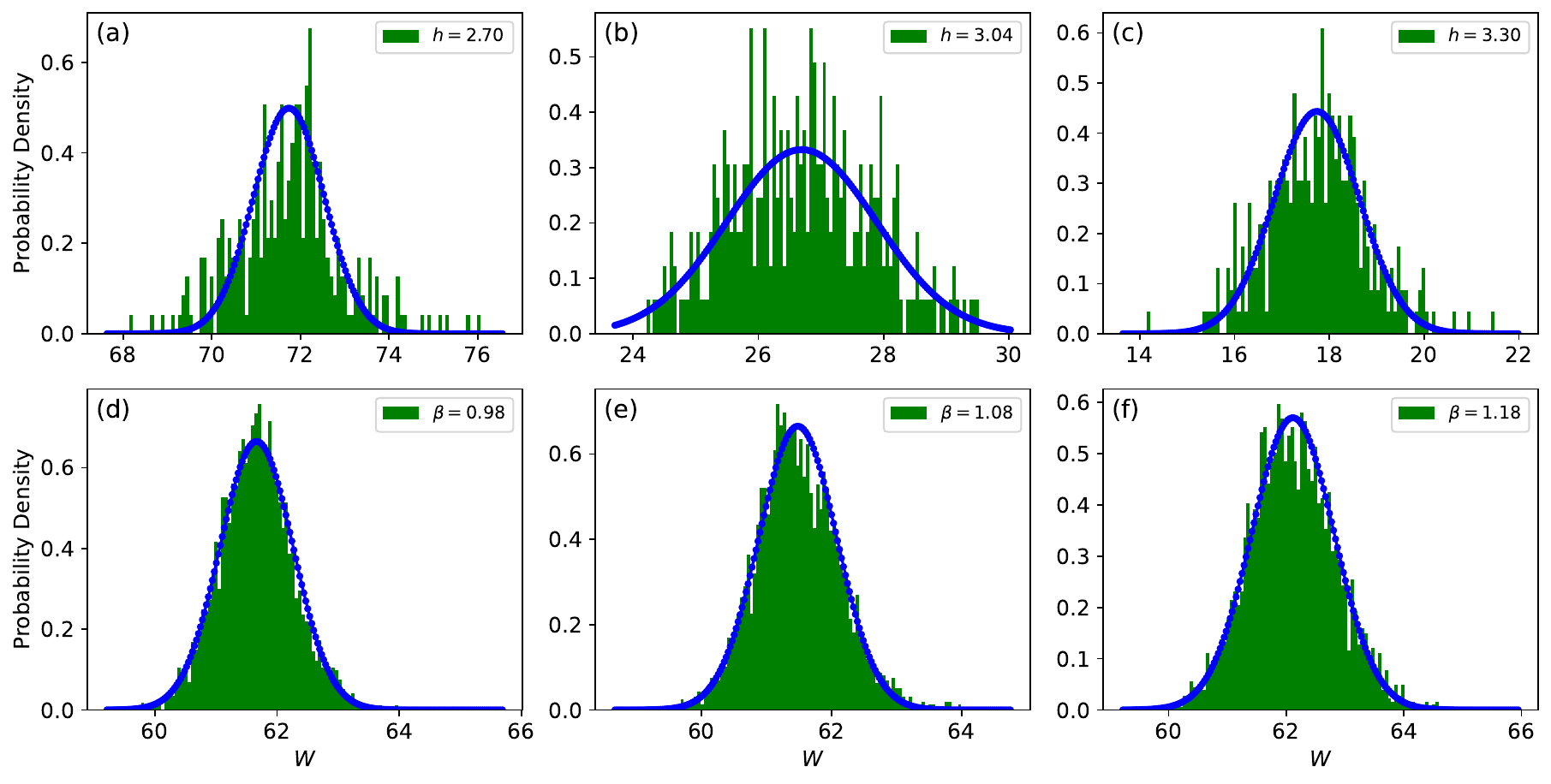}
\caption{Empirical probability distribution (green bars) of the measured work $\{W\}$ for (a)-(c) $T=0$ 2nd SRE, (d)-(f) $T>0$ 2nd SRE on $16\times16$ Ising model (PBC, $J=1$), with Gaussian distribution (blue lines) as fitting curves. } 
\label{fig:work_distr}
\end{figure*}

We begin by assuming that the sampled data ($\{W\}$) collected from the non-equilibrium QMC is described as 
\begin{equation}   
W=\overline{W}+\delta,\quad \delta\sim f(\delta)
\end{equation}
where  $f(\delta)=\frac{1}{\sqrt{2\pi\sigma^2}}e^{-\delta^2/2\sigma^2}$ with variance $\sigma^2$.  This assumption is consistent with the empirical distribution seen in Fig.~\ref{fig:work_distr}. We now will estimate from this the relative variance of the observable $e^{-W}$ .  We note that 
\begin{equation}
\mu\equiv\overline{e^{-W}}=e^{-\overline{W}}\int \text{d}\delta e^{-\delta}f(\delta)=e^{-\overline{W}}e^{\sigma^2/2}
\label{eq:expectSREfromWork}
\end{equation}
which subsequently gives the expectation of SRE as $M_n(\rho)=\frac{1}{1-n}\left(-\overline{W}+\sigma^2/2\right)<-\frac{1}{1-n}\overline{W}$. 
This is consistent with thermodynamic laws, as during this irreversible non-equilibrium process, some of the work will be dissipated into the environment, due to the increase of entropy, while the rest of the work contributes to the increase of free energy~\cite{Jarzynski1997}.   Next, we obtain the variance for $e^{-W}$ as below:
\begin{equation}
\overline{e^{-2W}}=e^{-2\overline{W}}\int \text{d}\delta e^{-2\delta}f(\delta)=e^{-2\overline{W}}e^{2\sigma^2}    
\end{equation}
\begin{equation}
\tau^2\equiv\overline{\left(e^{-W}-\overline{e^{-W}}\right)^2}=e^{-2\overline{W}}e^{\sigma^2}\left(e^{\sigma^2}-1\right)  
\label{eq:variance_exp_W}
\end{equation}
giving the relative standard deviation as $\left(\exp(\sigma^2)-1\right)^{1/2}$, which determines the cost for computing the free energy to a fixed error. We note that similar reasoning on the variance has also appeared in Ref.~\cite{ZhouMeng2024}.

\begin{figure}
\centering
\includegraphics[width=0.5\textwidth]{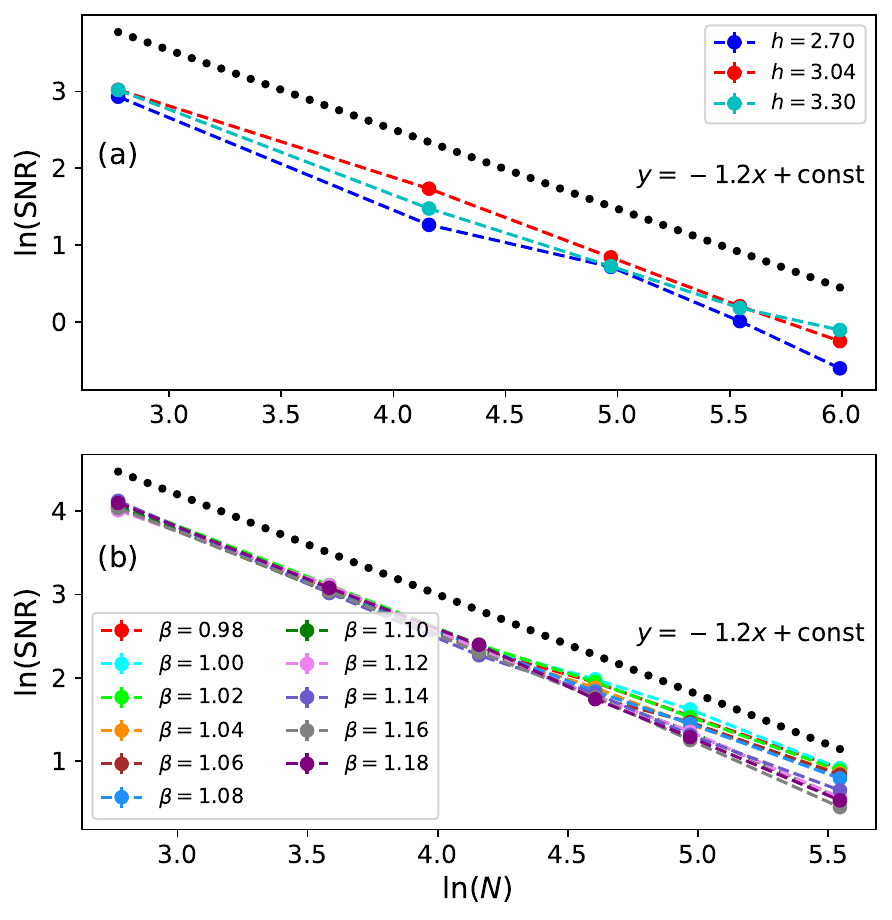}
\caption{Signal-to-noise ratio (SNR) versus system size $N$ on 2D $N=L^2$ transverse-field Ising model (PBC, $J=1$). (a) Ground state; (b) Finite-$T$ state at $h=2.75$. In both (a) and (b), straight lines $y=-1.2x+\text{const}$ are drawn for comparisons.}
\label{fig:sre_snr}
\end{figure}

From here, we will work out the variance of $W$  - i.e. $\sigma^2$
To do this, we turn to the numerical fitting for $\tau^2$, which is given by Fig.~\ref{fig:sre_snr}:
\begin{equation}
e^{\sigma^2}-1 = \alpha_c N^{\alpha}   
\end{equation}

\begin{figure*}
\centering
\includegraphics[width=1.0\textwidth]{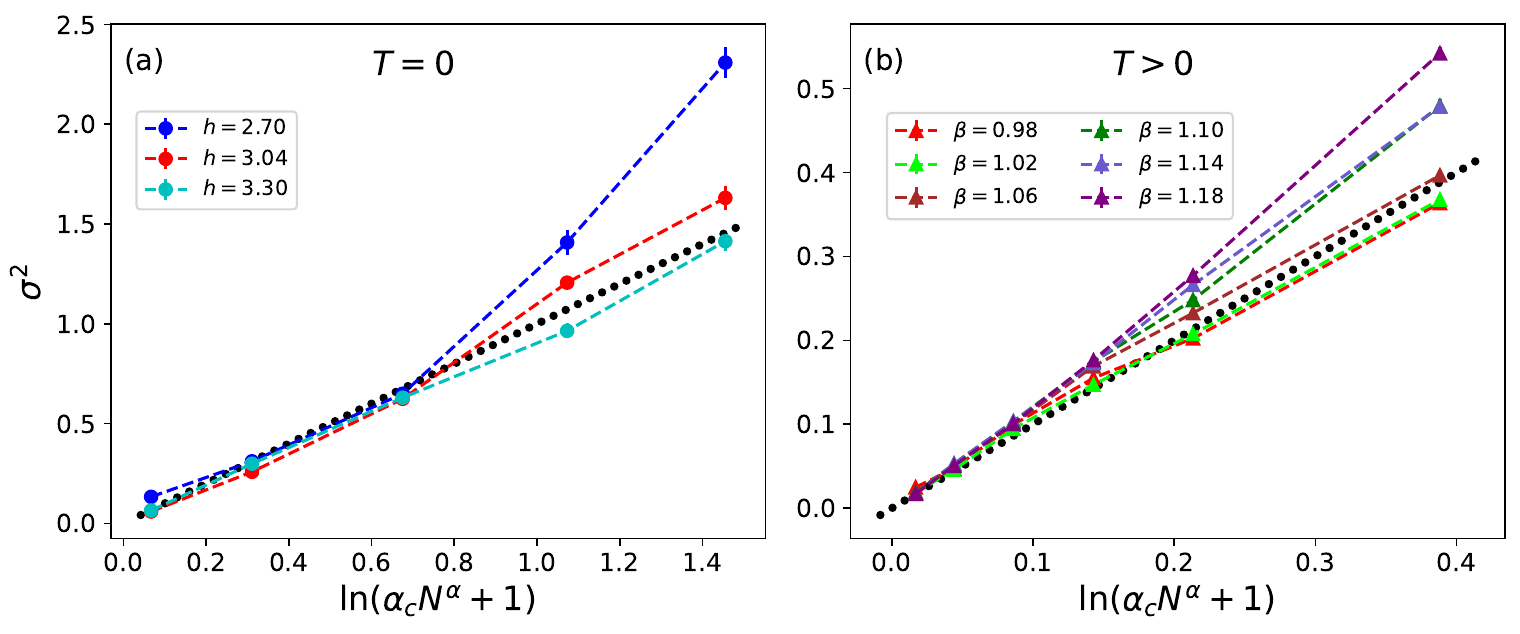}
\caption{Numerical examination on Eqn.~\ref{eq:sigma_N}, y-axis is the variance $\sigma^2$ directly evaluated from the sample set $W$, the fitting parameters $\alpha_c$ and $\alpha$ comes from the fitting on the SNR of $e^{-W}$ (a) $T=0$, $\alpha\simeq1.20$, $\alpha_c\simeq2.48\cdot10^{-3}$; (b) $T>0$, $\alpha\simeq1.20$, $\alpha_c\simeq6.11\cdot10^{-4}$. $y=x$ (black dotted lines) are drawn in (a) and (b) for comparisons.  } 
\label{fig:sigma_N}
\end{figure*}

Now we determine the constant factor $\alpha_c$ by fitting the SNR in Fig.~\ref{fig:sre_snr}. We have that
\begin{equation}
\sigma^2 = \ln\left( \alpha_cN^{\alpha}+1\right)
\label{eq:sigma_N}
\end{equation}
from which we expect that $\sigma^2\propto\ln N$ in the limit $N\to\infty$. We numerically verify Eqn.~\eqref{eq:sigma_N} in Fig.~\ref{fig:sigma_N}, where we compare the variance $\sigma^2$ directly from the sample set of $W$ with the scaling behavior using the fitting parameters from the SNR of $\{e^{W}\}$. We observe that except at $h=2.70$ (ferromagnetic phase), other comparisons agree well with our analysis towards Eqn.~\eqref{eq:sigma_N}. We suspect the disagreement in that case results from the deviation of the distribution of $W$ from our assumption of a Gaussian distribution on $W$ (see Fig.~\pref{fig:work_distr}{a}).  

\begin{figure*}
\centering
\includegraphics[width=0.9\textwidth]{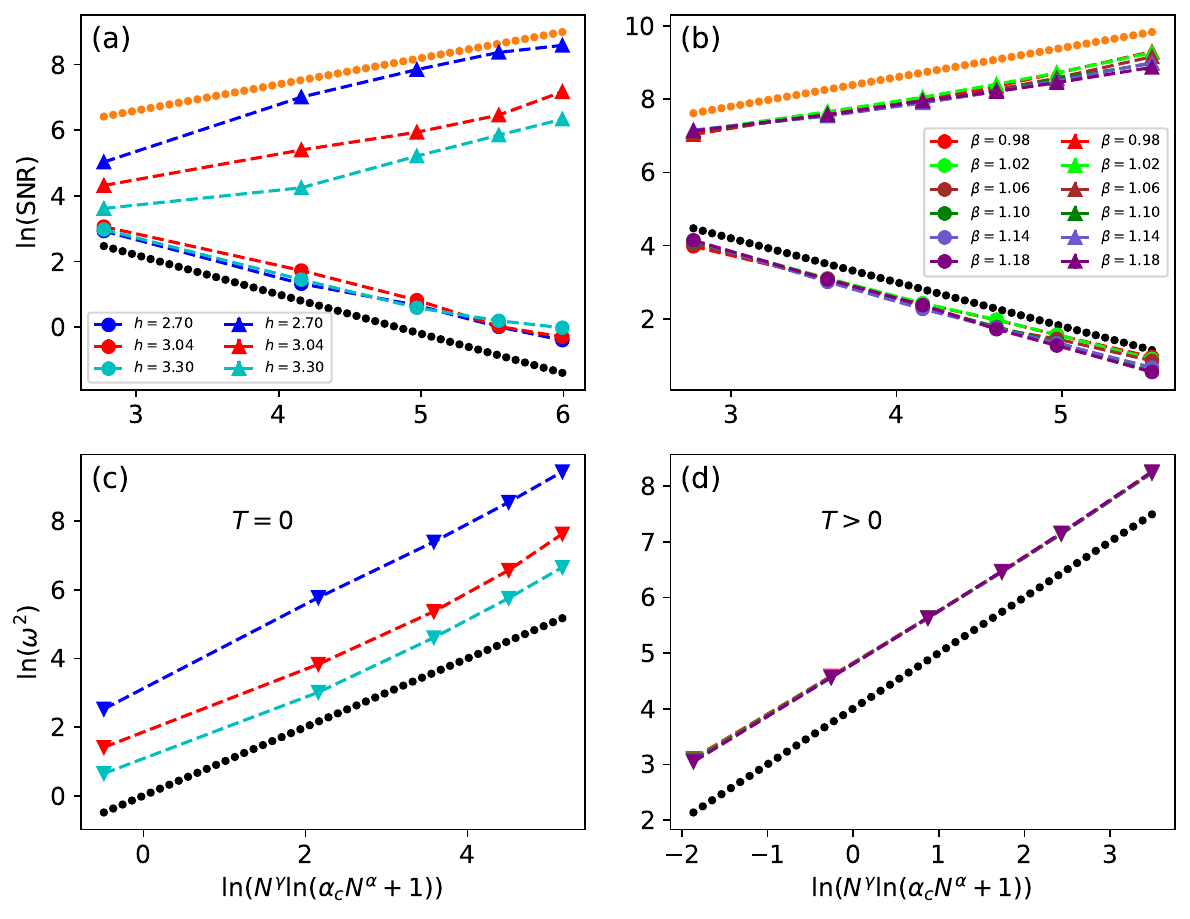}
\caption{Signal-to-noise ration (SNR) for $\{e^{-W}\}$ (circles) and $\{W\}$ (triangles) at (a) $T=0$ and (b) $T>0$ 2D Ising model (same parameters as Fig.~\ref{fig:work_distr}); Scaling behavior of $\{W\}$ w.r.t. system size $N$ at (c) $T=0$ and (d) $T>0$. Straight lines (dotted lines) are drawn for comparison in (a-b) with slope $\gamma\simeq0.80$ (orange) and in (c-d) slope=1 (black). } 
\label{fig:work_snr_suppl}
\end{figure*}

Furthermore, we can obtain the scaling of the mean of $W$ - i.e. $\omega$ - w.r.t. $N$. In Fig.~\pref{fig:work_snr_suppl}{a,b}, we add the SNR for the sample set of $W$, from which we fit the scaling as
\begin{equation}
\frac{\omega^2}{\sigma^2}\propto N^{\gamma}    
\end{equation}
where $\gamma\simeq0.80$ for both $T=0$ and $T>0$. Therefore, combined with Eqn.~\eqref{eq:sigma_N}, we will have
\begin{equation}
\omega^2 \propto N^{\gamma}\ln\left( \alpha_cN^{\alpha}+1\right)    
\label{eq:omega_N}
\end{equation}
and $\omega^2 \propto N^{\gamma}\ln N$ in the limit $N\to \infty$. Eqn.~\eqref{eq:omega_N} are examined numerically in Fig.~\pref{fig:work_snr_suppl}{c,d}, where we compare $\omega^2$ directly from the sample set of $W$ and the scaling with fitting parameters from SNR on $W$ and $e^{-W}$.

\section{A Unifying Framework for Exponential Observable Calculation}
{\label{append:framework}}

In this section, we show how our tensor network perspective applies directly to various other observables reproducing the known algorithm for using non-equilibrium QMC for the entanglement R\'enyi entropy (ERE) and generating a related algorithm for the entanglement R\'enyi negativity (ERN) and the Shannon R\'enyi negativity (or participation R\'enyi entropy, PRE).  
This view is related to the general approach of Ref.~\cite{ZhangPan2024} for exponential observables, i.e., $\ln\langle e^{\hat{X}}\rangle$ with an extensive operator $\hat{X}$, but instead uses the language of tensor networks and applies non-equilibrium QMC to contract them.  

Here we focus on the examples of ERE, ERN and PRE, which have played critical roles in characterizing the quantum correlation or localization within pure or mixed states. 

\begin{figure*}
\centering
\includegraphics[width=0.6\textwidth]{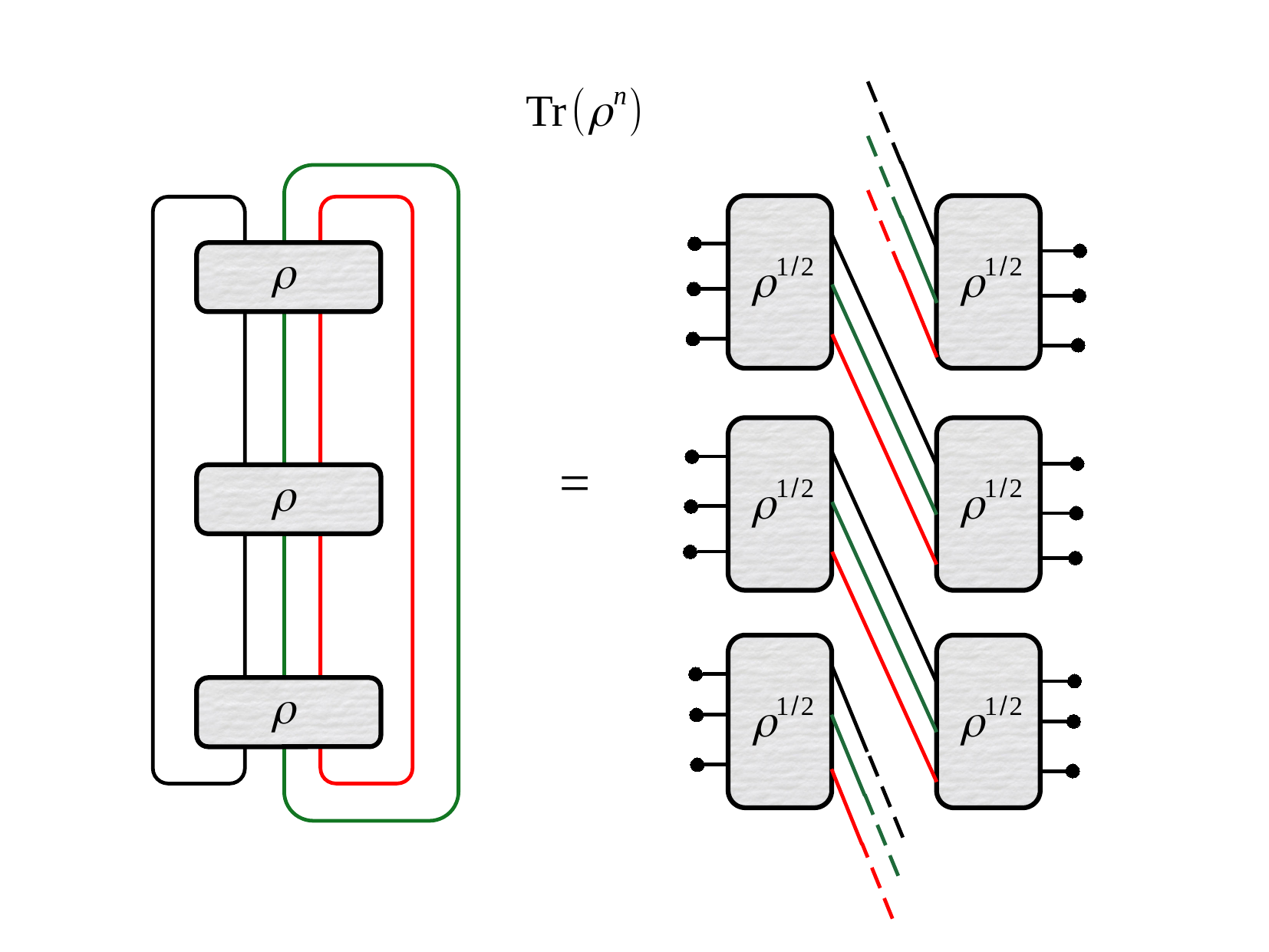}
\caption{Transformation of the tensor-network diagram for the numerate of ERE, i.e., $\Tr(\rho^n)$ ($n=3$ here), into a version analogous to Fig.~\ref{fig:keyPic}. Here we show the connections for three spin indices out of $\rho$, denoted with black, red, green lines, respectively.} 
\label{fig:ere}
\end{figure*}

The $n$-th ERE is defined as 
\begin{equation}
S_n=\frac{1}{1-n}\ln\frac{\Tr(\rho^n)}{(\Tr\rho)^n}    
\label{eq:ere_def}
\end{equation}
In general, $\rho$ comes from the trace over a subregion from a quantum system, but here, for simplicity, we will not consider this prior process. 

Now we will build up the connection to our framework by explicitly writing down the relevant tensors $T(B)$ and $\mathcal{T}$ for ERE. Firstly, we transform the tensor-network representation of the numerator from Eqn.~\eqref{eq:ere_def} into an analogous form to the one from Fig.~\ref{fig:keyPic}, as shown in Fig.~\ref{fig:ere}. From Fig.~\ref{fig:ere}, we can identify the $2n$-leg connection tensor $\mathcal{T}_S$ for $r$ replicas of $\rho$ as 
\begin{equation}
(\mathcal{T}_S)^{j_1,j_2,\cdots,j_r}_{k_1,k_2,\cdots,k_r}=\delta_{j_1,k_2}\delta_{j_2,k_3}\cdots\delta_{j_{r-1},k_r}\delta_{j_r,k_1}     
\end{equation}
where $\delta_{j,k}=1$ for $j=k$ and 0 otherwise. Consequently, we have
\begin{equation}
T_S(B)=\otimes_{s=1}^N\left(\mathcal{T}_S^{B_s}I_n^{1-B_s}\right)    
\end{equation}
for the generalized partition function
\begin{equation}
\mathcal{Z}_B=\Tr\left[\left(\rho^{1/2}\right)^{\otimes n}T_S(B)\left(\rho^{1/2}\right)^{\otimes n}\right]    
\label{eq:ZB_ere_def}
\end{equation}
where $\rho^{1/2}\rho^{1/2}\equiv\rho$. Note that here the spin set $B$ also plays the role of determining the presence of the connection tensor $\mathcal{T}_S$, and thus the connection topology for $\mathcal{Z}_B$, as done in Eqn.~\eqref{eq:def_TB}. 

\begin{figure*}
\centering
\includegraphics[width=0.6\textwidth]{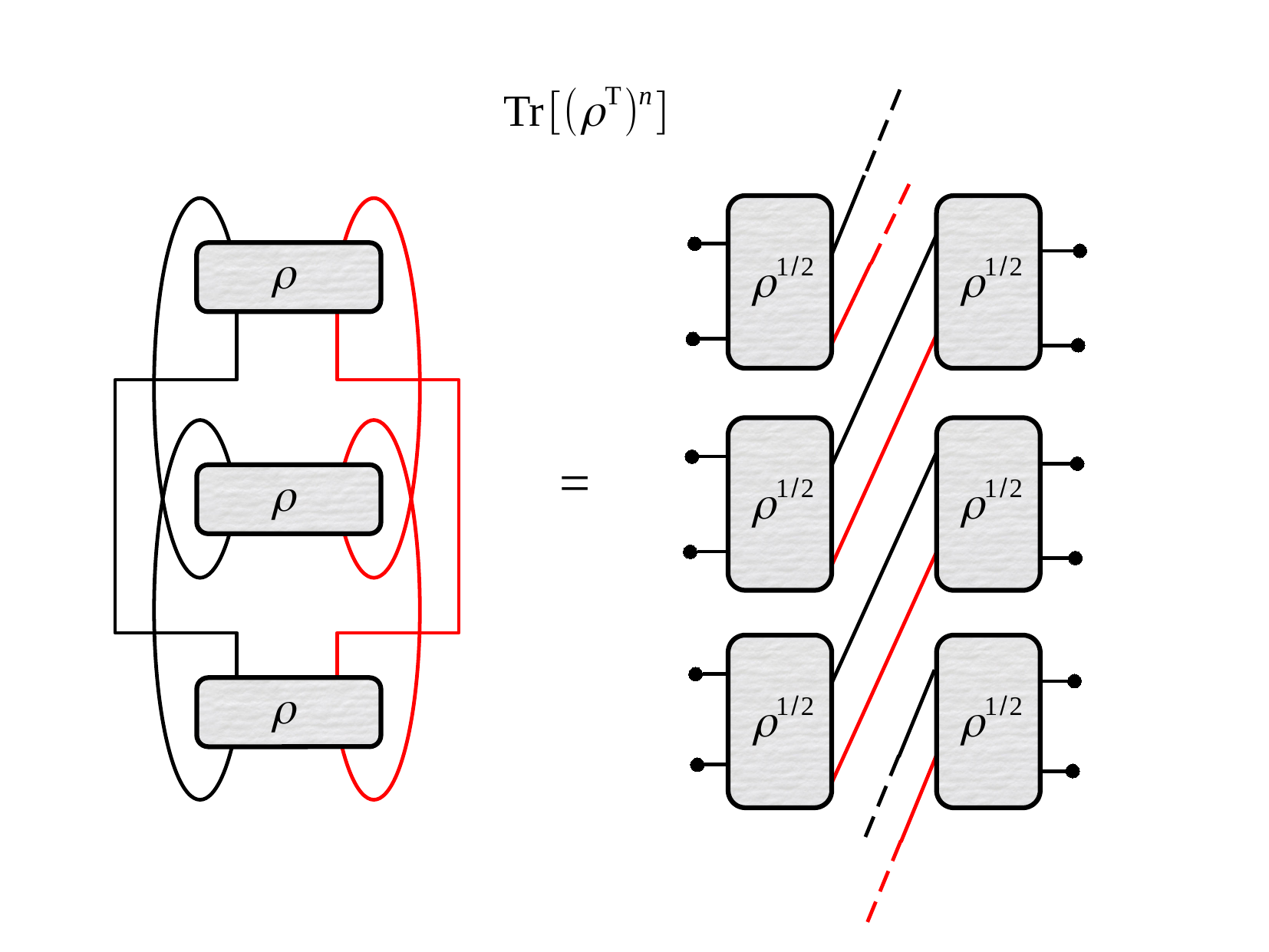}
\caption{Transformation of the tensor-network diagram for the numerate of ERN, i.e., $\Tr\left[(\rho^{T_A})^n\right]$ ($n=3$ here), into a version analogous to Fig.~\ref{fig:keyPic}. Here we only show the transpose on two spin sites, with the connections denoted by red and black lines respectively.} 
\label{fig:ern}
\end{figure*}

We write the definition for $n$-th ERN as 
\begin{equation}
G_n(A)=\frac{1}{1-n}\ln\frac{\Tr\left[(\rho^{T_A})^n\right]}{\Tr(\rho^n)}    
\label{eq:ern_def}
\end{equation}
where $\rho^{T_A}$ is obtained by performing the transpose on the subregion $A$ of the density matrix $\rho$. Note that the denominator of ERN is identical to the numerator of ERE that we have just discussed. Next, similarly, we identify the $2n$-leg connection tensor $\mathcal{T}_G$ from Fig.~\ref{fig:ern} as
\begin{equation}
(\mathcal{T}_G)^{j_1,j_2,\cdots,j_r}_{k_1,k_2,\cdots,k_r}=\delta_{j_1,k_r}\delta_{j_2,k_1}\cdots\delta_{j_{r-1},k_{r-2}}\delta_{j_r,k_{r-1}}         
\end{equation}
and the $2nN$-leg tensor $T_G(B)$ as
\begin{equation}
T_G(B)=\otimes_{s=1}^N\left(\mathcal{T}_G^{B_s}\mathcal{T}_S^{1-B_s}\right)    
\end{equation}
Here we have the constraint for the spin set $B$ as $B\subseteq A$ instead of $B\subseteq [N]$ from the SRE case.

Similarly, we define $n$-th PRE for a wave-function $\ket{\psi}$ and computational basis $\{j\}$ as 
\begin{equation}
\begin{split}    
H_n(\ket{\psi}) =& \frac{1}{1-n}\ln\sum_j\left(\braket{\psi}{j}\braket{j}{\psi}\right)^n\\
=& \frac{1}{1-n}\ln\frac{\bra{\psi}^{\otimes n}T_H([N])\ket{\psi}^{\otimes n}}{\bra{\psi}^{\otimes n}I^{\otimes n}\ket{\psi}^{\otimes n}}
\label{eq:pe_def}
\end{split}
\end{equation}
where we define the $2nN$-leg tensor $T_H(B)$ as 
\begin{equation}
T_H(B)=\otimes_{s=1}^N\left(\mathcal{T}_H^{B_s}I_n^{1-B_s}\right)    
\end{equation}
and the $2n$-leg tensor $\mathcal{T}_H$ as  
\begin{equation}
\mathcal{T}_H=\left(\ket{0}\bra{0}\right)^{\otimes n}+\left(\ket{1}\bra{1}\right)^{\otimes n}    
\end{equation}

With these tensor forms determined, the remaining algorithm is essentially identical. Explicitly, we first thermalize the configurations for $\mathcal{Z}_{\varnothing}$ in the main random walker. Next, we take the equilibrated configurations from the main walker as the start for the side random walkers, where we gradually increase the external parameter $\lambda$ from 0 to 1, and record the increments of $\Delta\ln g$ as the work $W$ during this process. At the end, the R\'enyi entropy, Eqn.~\eqref{eq:ere_def}, Eqn.~\eqref{eq:ern_def} or Eqn.~\eqref{eq:pe_def}, can be estimated as $\ln\langle e^{-W}\rangle$, following the Jarzynski’s equality~\cite{Jarzynski1997}.

\section{SRE of Mixed States}
{\label{append:mixSRE}}
In this work, we note that the SRE=0 states are the states generated by Clifford circuits, i.e., circuits made of Clifford gates and measurements; such a definition generalizes naturally from pure states to mixed states~\cite{Aaronson2004}. In this section, we will prove that the mixed states from such kind of Clifford circuits indeed have their 2nd SRE=0. 

To the beginning, we define an $N$-qubit stabilizer pure state $\ket{\phi}$ by the set of generators $\{g_1,g_2,\dots,g_N\}$ for its stabilizer group $S_{\ket{\phi}}=\langle g_1,g_2,\dots,g_N\rangle$, thus we have the density matrix as
\begin{equation}
\rho=\ket{\phi}\bra{\phi}=\frac{1}{2^N}\sum_{\bm{s}\in\{0,1\}^N}\prod_{j=1}^{N}g_j^{s_j} 
\end{equation}
Next, by measuring part of the pure state system and disregard the measurement outcomes, we obtain a mixed state
subsequently. We denote the left (measured) system part as $A$ ($\overline{A}$) with qubit number $N_A$ ($N_{\overline{A}}$). To proceed, we also rearrange the generators in such a way that the first $N_g$ of the generators $\{g_1,g_2,\dots,g_{N_g}\}$ only have identity operators on $\overline{A}$, while the rest have at least one non-identity operator on $\overline{A}$. Note that $0\leq N_g\leq N_A$, and  such rearrangements can be done by Gaussian elimination on the stabilizer tableau. Consequently, we have the density matrix for system $A$ as
\begin{equation}
\rho_A=\Tr_{\overline{A}}\rho=\frac{1}{2^{N_A}}\sum_{\bm{t}\in\{0,1\}^{N_g}}\prod_{j=1}^{N_g}g_j^{t_j}  
\end{equation}

Observing that 2nd ERE of $\rho_A$ is given by $S_2(\rho_A)=(N_A-N_g)\ln2$, $\rho_A$ is strictly a mixed state when $N_g<N_A$, but it is classically simulable with the generators $\{g_1,g_2,\dots,g_{N_g}\}$. Note that each generator $g_j$ has been traced over the region $\overline{A}$ such that it is composed of $N_A$ Pauli operators now. Finally, we will show that the 2nd SRE $M_2(\rho_A)=0$, which is done by showing the first term of Eqn~\eqref{eq:SRE_def} is exactly equal to $S_2(\rho_A)$: for any Pauli string $\sigma$,
\begin{equation}
\left[\Tr(\rho_A\sigma)\right]^2=
\begin{cases}
1, & \sigma\in\langle g_1,g_2,\dots,g_{N_g}\rangle \\
0, & \text{otherwise}
\end{cases}
\end{equation}
Henceforth, 
\begin{equation}
-\ln\sum_{\sigma\in\mathcal{P}_{N_A}}\frac{1}{2^{N_A}}\left[\Tr(\rho_A\sigma)\right]^4=(N_A-N_g)\ln 2=S_2(\rho_A)    
\end{equation}

At last, we investigate the properties of SRE when a mixed state cannot be represented with a single stabilizer tableau but can be decomposed as a series of stabilizer states. To this end, we consider a simple example in stabilizer code space from quantum error correction (QEC) ~\cite{nielsen2001quantum}. 

To be specific, let $\mathcal{C}$ be a $2^{N_A-N_g}$-dimensional code space for $N_A$ physical qubits, that holds $2^{N_A-N_g}$ quantum codes (mutually orthogonal stabilizer states) for $N_A-N_g$ logical qubits. These quantum codes share a common set of stabilizer generators $\langle g_1,g_2,\dots,g_{N_g}\rangle$, and are eigenstates of the $N_A-N_g$ logical-Z operators: $\{g_{N_g+1},g_{N_g+2},\dots,g_{N_A}\}$ with eigenvalues $\{\pm1\}$ (see Tab.~\ref{tab:qec}), so that their stabilizer tableau are completed by the addition of logical-Z operators with corresponding signs.

\begin{table}[ht]
\centering
\begin{tabular}{lc}
\toprule
\text{Quantum codes} & \text{Eigenvalue for logical-Z} \\
\midrule
$s_1$ & $1,1,\cdots,1$  \\
$s_2$ & $-1,1,\cdots,1$  \\
$s_3$ & $1,-1,\cdots,1$  \\
$\cdots$ & $\cdots$  \\
$s_{2^{N_A-N_g}}$ & $-1,-1,\cdots,-1$  \\
\bottomrule
\end{tabular}
\caption{Notations for the stabilizer quantum codes by the set of eigenvalues for logical-Z operators.}
\label{tab:qec}
\end{table}
Henceforth, one can consider the mixed states in this subspace $\mathcal{C}$ that can be represented as 
\begin{equation}
\rho=\sum_{k=1}^{2^{N_A-N_g}} a_k\ket{s_k}\bra{s_k},\quad \sum_k a_k=1   
\end{equation}
and the Pauli coefficients will only be nonzero when the Pauli strings belong to the stabilizer group:
\begin{equation}
\begin{split}    
&\left[\Tr(\rho\sigma)\right]^4=\left(\sum_{k=1}^{2^{N_A-N_g}}a_k\expect{s_k}{\sigma}{s_k}\right)^4\\
=&   
\begin{cases}
\left(\pm a_1\pm a_2\pm\cdots\pm a_{2^{N_A-N_g}}\right)^4, & \sigma\in\langle g_1,g_2,\dots,g_{N_A}\rangle \\
0, & \text{otherwise}
\end{cases}
\end{split}
\end{equation}
where the $\pm$ signs depend on the presence of logical-Z operators in $\sigma$. That is, among the nonzero Pauli coefficients, the ones without the logical-Z operators ($2^{N_g}$ such terms in total) are equal to $\left(a_1+a_2+\cdots+a_{2^{N_A-N_g}}\right)^4$, while the rest (with at least one logical-Z, $2^{N_A}-2^{N_g}$ terms in total) have half of signs positive and half negative inside the parentheses. 2nd SRE=0 for the following special cases: (1) $\rho$ is a pure stabilizer state, i.e., $a_k=1$ for a specific $k$, while the rest are 0; (2) $\rho$ is an equal mixture of all the quantum codes, i.e., $a_k=\frac{1}{2^{N_A-N_g}}$ for all $k$'s. In both cases, $\rho$ can be represented with one stabilizer tableau.


\end{document}